\documentclass[aps,prd,superscriptaddress,nofootinbib,10pt,english,tightenlines,floats,showpacs,showkeys]{revtex4}
\usepackage{amsmath,amstext,amssymb,amsfonts}
\usepackage{fontenc}
\usepackage{textcomp}
\usepackage{lipsum}

\usepackage{color}
\usepackage{graphicx}

\def\nn{\nonumber\\}
\newcommand{\f}[2]{\frac{#1}{#2}}
\def\be{\begin{equation}}
\def\ee{\end{equation}}
\def\bea{\begin{eqnarray}}
\def\eea{\end{eqnarray}}

\def\bwt{\begin{widetext}}
\def\ewt{\end{widetext}}

\DeclareMathOperator\arctanh{arctanh}
\begin{document}

\title{Einstein-Cartan gravitational collapse of a homogeneous Weyssenhoff fluid}

\author{Amir Hadi Ziaie}\email{ah_ziaie@sbu.ac.ir}
\affiliation{Department of Physics, Shahid Beheshti University, G. C., Evin,Tehran, 19839, Iran}
\author{Paulo Vargas Moniz}\email{pmoniz@ubi.pt}
\affiliation{Departamento de F\'{i}sica, Universidade da Beira Interior, Rua Marqu\^{e}s d'Avila
e Bolama, 6200 Covilh\~{a}, Portugal.}
\affiliation{Centro de Matem\'{a}tica e Aplica\c{c}\~{o}es (CMA - UBI),
Universidade da Beira Interior, Rua Marqu\^{e}s d'Avila
e Bolama, 6200 Covilh\~{a}, Portugal.}
\author{Arash Ranjbar}\email{a_ranjbar@sbu.ac.ir}
\affiliation{Department of Physics, Shahid Beheshti University, G. C., Evin,Tehran, 19839, Iran}
\affiliation{Centro de Estudios Cient´ıficos (CECs) Av. Arturo Prat 514, Valdivia, Chile}
\affiliation{Universidad Andr´es Bello, Av. Rep´ublica 440, Santiago, Chile}
\author{Hamid Reza Sepangi}\email{hr-sepangi@sbu.ac.ir}
\affiliation{Department of Physics, Shahid Beheshti University, G. C., Evin,Tehran, 19839, Iran}
\date{\today}
\preprint{hep-th/yymmnnn}

\begin{abstract}
We consider the gravitational collapse of a spherically symmetric homogeneous matter distribution consisting of a Weyssenhoff fluid in the presence of a negative cosmological constant. Our aim is to investigate the effects of torsion and spin averaged terms on the final outcome of the collapse.
For a specific interior spacetime setup, namely the homogeneous and isotropic FLRW metric,
we obtain two classes of solutions to the field equations where depending on the relation between spin source parameters, $(i)$ the collapse procedure culminates in a spacetime singularity or $(ii)$ it is replaced by a non-singular bounce. We show that, under certain conditions, for a specific subset of the former solutions, the formation of trapped surfaces is prevented and thus the resulted singularity could be naked. The curvature singularity that forms could be gravitationally strong in the sense of Tipler. Our numerical analysis for the latter solutions shows that the collapsing dynamical process experiences four phases, so that two of which occur at the pre-bounce and the other two at post-bounce regimes. We further observe that there can be found a minimum radius for the apparent horizon curve, such that the main outcome of which is that there exists an upper bound for the size of the collapsing body, below which no horizon forms throughout the whole scenario.
\end{abstract}
\maketitle

\section{Introduction}

The final state of the gravitational collapse of a massive star is one of the challenges in classical general relativity (GR) \cite{Joshi}. A significant contribution has been to show
that, under reasonable initial conditions, the spacetime
describing the collapse process would inevitably admit singularities \cite{HP}.
These singularities, can either
be hidden behind an event horizon\footnote{There is a recent discussion by Hawking \cite{Hawking-Info} arguing that this role is played instead by the apparent horizon, which is formed during the collapse process and is responsible to conceal the singularity from the outside observers.} or visible to distant observers.
In the former, a black hole forms
as the end product of a continual collapse process, as hypothesized by the cosmic
censorship conjecture\footnote{The CCC is categorized into two types, weak cosmic
censorship conjecture (WCCC) and the corresponding strong version (SCCC).
WCCC states that there can be no singularity communicating with asymptotic observers,
thus forbidding the occurrence of globally naked singularities, while SCCC asserts that timelike
singularities never occur, prohibiting the formation of locally naked singularities \cite{WSCCC}. Whereas the CCC is concerned with stability of solutions to Einstein\rq{}s field equations, there is a second class of censorship conjectures \cite{WCCCC} which asserts that all naked singularities are in some sense gravitationally weak.} (CCC) \cite{ccc} (See also \cite{RevonCCC} for reviews on the conjecture).
The latter are classified as naked singularities, whose existence in GR has been established under a variety of specific  circumstances and for different models, with matter content of various types, e.g. scalar fields \cite{G1,SF}, perfect fluids \cite{PF}, imperfect fluids \cite{IPF} and null strange quark fluids \cite{NSQF}. The analysis has also been taken to  wider gravitational settings, such as $f(R)$ theories \cite{Khodam}, Lovelock gravity \cite{Love} (see also \cite{nakedrecentr} for some recent reviews) and hypothesized quantum gravity theories \cite{LQGSR,WEAKVIO,LQGSRCS}. This is an interesting line of research because, the possible discovery of naked singularities may provide us with an opportunity to extract  information from physics beyond trans-Planckian regimes \cite{transplanck}, see e.g. (\cite{ODNS} for the possibility of observationally detecting naked singularities).\label{ff}

It is therefore well motivated to consider other realistic gravitational theories whose geometrical attributes (not present in GR) may affect the final asymptotic stages of the collapse.  More concretely, could ingredients mimicking spin effects
(associated with fermions) potentially influence the final fate of a collapse scenario? In
fact, if spin effects are explicitly present then GR will no longer be the theory to
describe the collapse dynamics. In GR, the energy-momentum couples to the
metric. However, when the spin of particles is introduced into the
framework, it is expected to couple to a geometrical quantity related to the rotational degrees
of freedom in the spacetime. This point of view suggests a spacetime which is non-Riemannian, namely generalizations of GR induced from the explicit presence of matter with such spin degrees of freedom \cite{Freedman:2012zz,Ortin:2004ms,HehlB}. One such framework, which will allow non-trivial dynamical consequences to be extracted is
the Einstein-Cartan (EC) theory \cite{HehlB,Hehl} where the metric and
torsion determine the geometrical structure of spacetime\footnote{Somewhat
related to such settings, let us mention the teleparallel theories
of gravity \cite{TEL}, as well as the Lyra theory \cite{Lyra}.
The latter concerning the propagation of torsion by means of
expressing it as the gradient of a scalar field. This scalar field can be suitably regarded as a gauge
function, describing a torsion potential or as a
Brans-Dicke scalar field which couples non-minimally to
curvature \cite{Soleng}.}. The torsion
can be interpreted as caused by  microscopic effects, e.g., by
 fermionic fields which are not taken explicitly into
account \cite{Ortin:2004ms}.

Within the  context of EC theories, it has been shown that considering the induced repulsive effects extracted from (averaged) spin interactions, the Big-Bang singularity can be
replaced by a non-singular bounce, before which the universe was contracting and halts at a minimum but finite radius \cite{GAS}. However, a curvature singularity as
the final fate of a gravitational collapse process can still occur even if
explicit spin-torsion and spin-spin repulsive
interactions \cite{match} are taken into account. The argument
that has been put forward is that since photons neither produce nor interact with the spacetime
torsion, the causal structure of an EC manifold, determined
by light signals, is the same as in GR. Hence the singularity theorems
in GR can be generalized to the EC theory by taking into account a
combined energy-momentum tensor which would include, by means of
some suitable averaging procedure, spin contributions
\cite{Hehl-Heyde}.

The results conveyed within this paper are twofold. We consider a
spherically symmetric configuration in the presence of a negative cosmological constant \cite{CC, DyCCH} whose matter
content is assumed to be a homogeneous Weyssenhoff fluid \cite{Weys} that collapses under its own gravity.
On the one hand, the first class of our solutions is shown to
evolve towards a spacetime singularity where the role of the negative cosmological constant is to set up the gravitational attraction through a positive pressure term. Then, as the collapse proceeds, a repulsive pressure computed from averaged spin-spin and spin-torsion interactions, balances the inward pressure, preventing trapped surfaces from forming in the later stages. Thus, the resulting singularity could be at least locally naked. Moreover, it is pertinent to point out that our analysis shows that, depending on the spin and  energy density
parameters, trapped surfaces can either be formed or avoided. On the other hand, second class of solutions suggest that the spin contributions to the field equations may generate a bounce that averts the formation of a spacetime singularity. Let us
furthermore note that, in contrast to some alternative theories of
gravity e.g., the Gauss-Bonnet theory  in which the
Misner-Sharp energy is modified \cite{GB}, our approach will involve only the manipulation of the
matter content. Hence the Misner-Sharp energy which is the key
factor that determines the dynamics of the apparent horizon is
defined in the same manner  as that in GR \cite{Hay}.

The organization of this paper is then as follows. In section (\ref{EQM}) we
present a brief review on the background field equations of the EC
theory in the presence of a Weyssenhoff fluid and a negative cosmological constant. Section (\ref{SOL})
provides a family of solutions, some of which
represent a collapse scenario that leads to a spacetime singularity
within a finite amount of time. In section (\ref{DAH}) we study the dynamics of
apparent horizon and induced spin effects on the formation  of trapped
surfaces and  show that trapped surface avoidance can occur for a subset of collapse settings. We examine the curvature strength of the naked singularity in  \ref{Strength} and show that the singularity is gravitationally strong in the sense of Tipler \cite{Tipler}. A second class of solutions exhibiting a bounce is presented in  (\ref{NSS}) where we show how the presence of a spin fluid could affect the dynamics of the apparent horizon. In section (\ref{Exterior}) we present a suitable solution for an exterior region and  discuss therein the matching between interior and exterior regions. Finally, conclusions are drawn in section (\ref{CON}).

\section{Equations of Motion}\label{EQM}

The action for the EC theory can be written in the form \cite{Ortin:2004ms}
\bea S=\int  d^4x\sqrt{-g}
\left\{\f{-1}{\kappa^2}(\hat{R}+2\Lambda)+\mathcal{L}_m\right\}
=\int d^4x \sqrt{-g}\bigg\{\f{-1}{\kappa^2}\bigg[R(\{\})+K^{\alpha}\!\!~_{\rho\alpha}K^{\rho\lambda}\!\!~_{\lambda}
K^{\alpha}\!\!~_{\rho\lambda}K^{\rho\lambda}\!\!~_{\alpha}+2\Lambda\bigg]+{\mathcal L}_m\bigg\}, \label{action}
\eea
 where $\hat{R}$ is the Ricci scalar constructed from the general affine connection $\hat{\Gamma}^{\alpha}_{~\mu\nu}$ and can be expressed, in general, as a function of independent background fields, i.e., the metric field $g_{\mu\nu}$ and the affine connection. The quantity $K_{\mu\nu\alpha}$ is the contorsion tensor defined as
\be\label{contortion}
K^{\mu}_{~\alpha\beta}=\f{1}{2}\left(T^{\mu}_{~\alpha\beta}-T_{\alpha~\beta}^{~\mu}-
T_{\beta~\alpha}^{~\mu}\right),\ee with the spacetime torsion tensor $T^{\alpha}_{~\mu\nu}$ being
geometrically defined as the antisymmetric part of the general affine
connection
\be\label{TEQ}
T^{\alpha}_{~\mu\nu}=\hat{\Gamma}^{\alpha}_{~\mu\nu}-\hat{\Gamma}^{\alpha}_{~\nu\mu}.
\ee
and $\mathcal{L}_m$ is the matter Lagrangian; $\Lambda$ is the cosmological constant.
We take the metric signature\footnote{For the sake of generality, we keep $\kappa$ and $\Lambda$ throughout the equations but in plotting the diagrams we set the units so that $c=\hbar=\kappa=1$ and $\Lambda=-1$.} as $(+,-,-,-)$ and $\kappa^2\equiv16\pi
G$. The presence of torsion in the macroscopic structure of spacetime
can, theoretically, be attributed to microscopic
fermionic matter fields with spin-angular momentum degrees of
freedom. In this paper, we take the matter part of the action to be
described by  a Weyssenhoff fluid \cite{Weys} which macroscopically
is a continuous medium but also conveys features  that can be
suitably associated with the (averaged) spinor degrees of
freedom of microscopic matter fields. Moreover, it has been shown
that with the assumption of the Frenkel condition (also known as
the Weyssenhoff condition), the setup may  be equivalently described
by means of an effective fluid in a plain GR setting where the
effective energy momentum tensor contains additional (spin induced)
terms \cite{GEA}. More concretely, the Lagrangian for the matter
content can be subsequently decomposed as \be \mathcal{L}_m=\mathcal{L}_{{\rm SF}}+\mathcal{L}_{{\rm AC}}.
\ee
The Lagrangian $\mathcal{L}_{{\rm SF}}$ contains the induced effects of a spinning fluid which
can be written in terms of a perfect fluid contribution and a
characteristic spin part \cite{Gasprini}\footnote{This term can be represented by an effective four fermion interaction which, together with a part from a Dirac Lagrangian, can be realized as Nambu-Jona-Lasinio effective action in $4$D.}. $\mathcal{L}_{{\rm AC}}$ conveys a minimal coupling of a spinor axial current with torsion \cite{Shapiro}\footnote{In more detail, $\mathcal{L}_{{\rm AC}}$ can be associated to a chiral interaction that corresponds to the coupling of contorsion to the massless fermion fields due to a massless Dirac Lagrangian in a curved background.}. Therefore, we may write
\be\label{axial}
\mathcal{L}_{{\rm AC}}=J^{\mu}S_{\mu},
\ee
where $J^{\mu}=\langle\bar{\psi}\gamma^5\gamma^{\mu}\psi\rangle$ and
$S_{\mu}=\epsilon_{\alpha\beta\rho\mu}T^{\alpha\beta\rho}$ correspond to the (averaged) spinor axial current\footnote{The axial current has been pointed out in the literature as responsible
for the Lorentz violation. Constraints have been imposed  on some of the
torsion components due to recent Lorentz violation investigations \cite{kostel}.
} and axial torsion vector, respectively\footnote{$\gamma^{\mu}$ is defined by
$\left[\gamma^{\mu},\gamma^{\nu}\right]=2g^{\mu\nu}$,
$\gamma^5=\f{-i}{\sqrt{-g}}\gamma_{0}\gamma_{1}\gamma_{2}\gamma_{3}$ is a chiral
Dirac matrix, $\psi$ is a fermion
field, $\bar{\psi}=\psi^{\dagger}\gamma^{0}$ is the conjugate
fermion field and $\epsilon_{\alpha\beta\rho\mu}$ is the totally
antisymmetric Levi-Civita tensor.}. Varying the action with
respect to the contorsion together with using (\ref{contortion}) gives \cite{GEA}
\be\label{constraint}
T^{\mu\nu\alpha}+2g^{\mu[\nu}T^{\alpha]}=\kappa^2
\tau^{\alpha\nu\mu}, \ee where $T^{\mu}=T^{\rho\mu}_{~~~\rho}$ and
$\tau^{\mu\nu\alpha}$ is the spin angular momentum tensor given by \be
\tau^{\mu\nu\alpha}:=\f{1}{\sqrt{-g}}\f{\delta(\sqrt{-g}\mathcal{L}_m)}{\delta
K_{\mu\nu\alpha}}:=\tau^{\mu\nu\alpha}_{{\rm AC}}+\tau^{\mu\nu\alpha}_{{\rm SF}}.
\ee
Moreover, considering the decomposition of the spin angular momentum  as $\tau^{\mu\nu\alpha}_{{\rm SF}}=-\f{1}{2}S^{\mu\nu}u^{\alpha}$ \cite{Weys}, where
$u^{\alpha}$ is the fluid 4-velocity,  $S^{\mu\nu}$ is  the
antisymmetric spin density tensor  representing the effective source
of torsion. In the EC theory, in contrast to the metric, the torsion is not
really a dynamical field; the left hand side of equation
(\ref{constraint}) contains no derivatives of the torsion tensor and
indeed appears as a purely algebraic equation.  Torsion can
therefore be eliminated by replacing it with the spin density
$S^\mu_{~\nu}$  and hence implying  a modification to
the Einstein field equations. Using the Frenkel
condition\footnote{This translates as saying that the intrinsic spin
contribution (in the form {of} the antisymmetric spin density
tensor) of a matter field is spacelike in the rest frame of the
fluid.}, $S^{\mu\nu}u_{\nu}=0$, the torsion constraint equation
(\ref{constraint}) may be rewritten in the form
\be\label{Tconstraint}
T^{\mu\nu\alpha}=-\kappa^2\left(2\epsilon^{\mu\nu\alpha\rho}J_{\rho}-\f{1}{2}S^{\nu\alpha}u^{\mu}\right).
\ee
At this point it is useful to show how the  Weyssenhoff fluid can fit the
Frenkel condition. More precisely, this condition results in an algebraic
relationship between the spin density tensor and
torsion as \bea T_{\nu}=T^{\mu}_{~\nu\mu}=\kappa u^{\mu}S_{\nu\mu},
\eea which can also be retrieved directly from the formalism
proposed in \cite{obuk}. Therefore, by virtue of the Frenkel condition, this means that the only
remaining degrees of freedom of the torsion are the traceless components of the torsion tensor.
Furthermore, the axial torsion vector can be written, with the
assistance of equations (\ref{contortion}) and (\ref{Tconstraint}),
as \be\label{axialvector}
S_{\mu}=\epsilon_{\alpha\beta\rho\mu}\left(2K^{\alpha\beta\rho}\right)=12\kappa^2
J_{\mu}+\f{1}{2}\kappa^2\epsilon_{\alpha\beta\rho\mu}S^{\alpha\beta}u^{\rho}.
\ee
It is now straightforward to obtain the dynamical equations of
motion, varying the action (\ref{action}) with respect to the
dynamical field $g^{\mu\nu}$ which can be written as \bea\label{geq}
G_{\mu\nu}&-&\Lambda g_{\mu\nu}-K^{\alpha}_{~\mu\alpha}K^{\lambda}_{~\nu\lambda}-\f{1}{2} T^{\alpha}_{~\rho\mu}T^{\rho}_{~\nu\alpha}-\f{1}{2}T^{\alpha}_{~\mu\lambda}T_{\alpha~~\nu}^{~\lambda}
-\f{1}{4}T_{\mu\rho\lambda}T_{\nu}^{~\rho\lambda}\nn
&+&\f{1}{8}g_{\mu\nu}\left(4K^{\alpha}_{~\rho\alpha}K^{\lambda\rho}_{~~\lambda}+2T^{\alpha\rho\lambda}T_{\rho\lambda\alpha}-T^{\alpha\rho\lambda}T_{\alpha\rho\lambda}\right)
=\f{\kappa^2}{2}T_{\mu\nu}. \eea
Substituting  for the contorsion from equation (\ref{contortion}) into the above equation
and using equation (\ref{Tconstraint}), we subsequently get
\bea\label{EOM1}
G_{\mu\nu}-\Lambda g_{\mu\nu}&-&\kappa^4\bigg[g_{\mu\nu}J^2+2J_{\mu}J_{\nu}-\f{1}{4}g_{\mu\nu}J_{\sigma}
\epsilon^{\alpha\rho\lambda\sigma}S_{\rho\lambda}u_{\alpha}-\f{1}{2}J_{\sigma}
\epsilon_{(\mu}^{~~\rho\lambda\sigma}
u_{\nu)}S_{\rho\lambda}\nn
&+&\f{1}{32}g_{\mu\nu}S_{\rho\lambda}S^{\rho\lambda}-\f{1}{8}S_{\mu\lambda}S_{\nu}^{~\lambda}+
\f{1}{16}u_{\mu}u_{\nu}S_{\rho\lambda}S^{\rho\lambda}\bigg]=\f{\kappa^2}{2}\big(T_{\mu\nu}^{{\rm SF}}+T_{\mu\nu}^{{\rm AC}}\big),
\eea
where $J^2=J_{\mu}J^{\mu}$. The energy momentum tensor contains
two contributions from the axial current, $T_{\mu\nu}^{{\rm AC}}$, and
the spin fluid, $T_{\mu\nu}^{{\rm SF}}$, which can be expressed\footnote{The spin-spin and
spin-torsion interactions are only significant over microscopic ranges, i.e.,
at sufficiently high matter densities. This means that the EC theory does
not directly challenge general relativity at large scales.
In order to take into account the macroscopic effects of spin contributions within the
framework of EC theory, a suitable averaging of the spin is assumed \cite{Nurga}. It is worth mentioning that in the process of taking the
average of a spherically symmetric isotropic system of randomly oriented spin particles, the average of the spin density
tensor is assumed to vanish, $\langle S^{\mu\nu}\rangle$=0,  but for the spin squared terms $\langle S^{\mu\nu}S_{\mu\nu}\rangle\neq0$.} after
employing a suitable spin averaging as \bea
\langle T^{{\rm AC}}_{\mu\nu}\rangle&=&-8\kappa^2 g_{\mu\nu} J^2-4\kappa^2 J_{\mu}J_{\nu},\label{Tac}\\
\langle
T^{{\rm SF}}_{\mu\nu}\rangle&=&-\f{2\kappa^2}{3}u_{\mu}u_{\nu}\sigma^2+\f{\kappa^2}{6}g_{\mu\nu}\sigma^2
+\left[(\rho_{{\rm SF}}+p_{{\rm SF}})u_{\mu}u_{\nu}-p_{{\rm SF}}
g_{\mu\nu}\right],\label{Tsf} \eea
where we have replaced the various spin-averaged quantities with \cite{GEA}
\bea
\langle S_{\mu\nu}S^{\mu\nu}\rangle&=&2\sigma^2,\\
\langle S_{\mu}^{~~\rho}S_{\nu\rho}\rangle&=&\f{2}{3}\left(g_{\mu\nu}-u_{\mu}u_{\nu}\right)\sigma^2,\\
\langle S_{\mu\nu} J^{\alpha}\rangle&=&0. \eea
From a macroscopic point of view, the spin fluid can therefore be considered as a
contribution from a conventional  perfect fluid  with the associated energy
density $\rho_{{\rm SF}}$ and pressure $p_{{\rm SF}}$ plus the first two terms in equation (\ref{Tsf}), which represent characteristic spin contributions and arise from a suitably averaged microscopic treatment of the fluid. Inserting equations (\ref{Tac}) and (\ref{Tsf}) into (\ref{EOM1}), together with the above spin averaging, we finally obtain the dynamical
field equations  as \cite{GEA}
\bea\label{FEs}
G_{\mu\nu}-\Lambda g_{\mu\nu}&=&\kappa^4\left[-3g_{\mu\nu}J^2+\f{1}{16}g_{\mu\nu}\sigma^2-\f{1}{8}u_{\mu}u_{\nu}\sigma^2\right]
+\f{\kappa^2}{2}\left[(\rho_{{\rm SF}}+p_{{\rm SF}})u_{\mu}u_{\nu}-p_{{\rm SF}}g_{\mu\nu}\right].
\eea

\section{Solutions to the field equations}\label{SOL}

In this section, we will find a class of collapse solutions which lead to the formation of a spacetime singularity. If the spacetime is assumed to have fewer symmetries (that is, inhomogeneities or anisotropies), there is a paucity of physically reasonable exact solutions available owing to the intrinsic difficulties. We therefore restrict the discussion  to a homogeneous and isotropic interior line element, representing the FLRW geometry  \cite{HP}
\be\label{metric}
ds^2=dt^2-a^2(t)dr^2-R^2(t,r)d\Omega^2,
\ee
where $R(t,r)=ra(t)$ is the physical radius of the collapsing matter, with
$a(t)$ being the scale factor and $d\Omega^2$  the standard line element
on the unit 2-sphere. The field equations for the above metric read
\bea
3H^2-\Lambda&=&\kappa^4\left[-3J(t)^2-\f{\sigma(t)^2}{16}\right]
+\f{\kappa^2}{2}\rho_{{\rm SF}}(t)\equiv \rho_{{\rm eff}}(t),\label{FEs00}\\
-2\dot{H}-3H^2+\Lambda&=&\kappa^4\left[3J(t)^2-\f{\sigma(t)^2}{16}\right]
+\f{\kappa^2}{2}p_{{\rm SF}}(t)\equiv
p_{{\rm eff}}(t),\label{FEs11} \eea
where $H=\dot{R}/R=\dot{a}/a$
is the rate of collapse.  Since we are interested in a continual
collapse process, $\dot{R}(t,r)$ must be negative. Notice that we may consider the cosmological constant term as vacuum energy density \cite{Carroll}. The continuity
equation for the matter fluid is therefore \bea\label{CE}
\dot{\rho}_{{\rm SF}}+3H(\rho_{{\rm SF}}+p_{{\rm SF}})=\kappa^2\left[12J\dot{J}+\frac{\sigma\dot{\sigma}}{4}+\f{3H}{4}\sigma^2\right].
\eea

As we mentioned in section \ref{EQM}, the geometry and matter content
effectively induce a macroscopic perfect fluid contribution with the
barotropic equation of state $p_{{\rm SF}}=w \rho_{{\rm SF}}$, together with
intrinsic spin contributions that are present  in the form of
averaged quadratic terms which may admit a
possible microscopic representation as, e.g., unpolarized fermions\footnote{In a collapse setting whose matter content is explicitly fermion dominated it is
conceivable that the effective spinning fluid might be
polarized. Thus, a spin alignment due to the presence of strong magnetic fields
 (cf. \cite{SPol}) may potentially affect the collapse dynamics and
therefore, quite possibly, its final outcome. Moreover, from a macroscopic viewpoint, each particle
in the cluster undergoing gravitational collapse may also have orbital angular momentum, so that the net effect of all the particles is to introduce a nonzero tangential pressure in the energy momentum tensor. Such rotational effects on the collapse process (e.g., gravitational collapse of a system of counter-rotating particles - the\lq\lq{} Einstein cluster\rq\rq{}\cite{ECluster}) have been studied in \cite{Counter}. It is shown there that trapped surface formation can be avoided, and so the singularity can be visible, if the angular momentum is strong enough.\label{f7}}. It is plausible to assume that the fermions participating in the collapse
process behave as ultra-relativistic particles. Thus the number
density ($n_{\rm f}$) of a fermionic gas, satisfying the Fermi-Dirac distribution,
can be approximated by $n_{\rm f}\propto a^{-3}$, see \cite{PADMAN} for more details. So the squares of the spin density and the axial current, which are
proportional to $ n_{\rm f}^2$ \cite{Gasprini,GEA,Nurga}, depend on the scale
factor as\footnote{In general, $\sigma_0^2$ and $J_0^2$ are the
source parameters for the squared spin density and axial current
respectively, and should not be confused with their initial
values defined as $\sigma_i^2=\sigma_0^2 a_i^n$ and $J_i^2=J_0^2 a_i^n$.}
\be\label{Js}
J^2=J_0^2 a^{-6},~~~~~ \sigma^2=\sigma_0^2 a^{-6}.
\ee
However, we proceed with a general setup \cite{Kuchow} where $J^2=J_0^2 a^n$ and $\sigma^2=\sigma_0^2 a^n$ ($n\in \mathbb{R}^{-}$).
Therefore, from (\ref{CE}) it is easy to obtain\footnote{The
choice $w=-1$ is discussed separately at the end of subsection \ref{NSS},
because it corresponds to a non-singular case.} the energy density
\bea\label{rhonew}
\rho_{{\rm SF}}(a)&=&C_0 a^{-3(1+w)}+\f{\alpha}{n+3(1+w)} a^{n},
\eea
where
\bea\label{c0al}
C_0&=&\left(\rho_{i_{{\rm SF}}}-\f{\alpha a_i^n}{n+3(1+w)}\right)a_i^{3(1+w)},\nn
\alpha&=&\f{\kappa^2}{8}\left[(n+6)\sigma_0^2+48nJ_0^2\right],
\eea
and we take $n+3(1+w)\neq 0,  w\neq-1$. Solving  equations (\ref{FEs00}) and (\ref{FEs11}) with the use of (\ref{rhonew}) leads to
\bea
\left(\f{\dot{a}}{a}\right)^2&=&\ell a^n+\f{\kappa^2}{6}C_0 a^{-3(1+w)}+\f{\Lambda}{3},\label{HH}\\
2\f{\ddot{a}}{a}&=&\ell(n+2)a^n-\f{\kappa^2}{6}C_0(1+3w)a^{-3(1+w)}+\f{2\Lambda}{3},\label{addot}\nn
\eea
where $\rho_{i_{{\rm SF}}}$ and $a_{i}$ are the initial values of the energy density and  scale
factor respectively, and
\be\label{ELL}
\ell=\kappa^4\left[\f{\sigma_0^2 (1-w)-48J_0^2(1+w)}{16(n+3(1+w))}\right].
\ee
From equation (\ref{HH}), in order to calculate the singularity time we need to evaluate the following integral
\bea\label{INT}
t_i\!\!\!-t=\int_{a_i}^{a(t)}\frac{da}{a\left[\ell a^n+\frac{\kappa^2}{6}C_0a^{-3(1+w)}+\frac{\Lambda}{3}\right]^{\frac{1}{2}}}
=\int_{a_i}^{a(t)}\f{da}{\sqrt{\ell}}\left[1+\frac{\kappa^2C_0}{6\ell}a^{-(n+3(1+w))}+\frac{\Lambda}{3\ell}a^{-n}\right]^{-\frac{1}{2}}a^{-\frac{n+2}{2}}.
\eea

The initial physical radius of the collapsing volume, $R_0=a_ir$, can be chosen so that the scale factor starts at $a_i$ and as the collapse proceeds the scale factor decreases ($\dot{a}<0$). Therefore, following Theorem 3.1 in~\cite{G1}, if we take the interval ($0,a_i$) to be sufficiently small we can use the binomial
expansion to evaluate the term under the square root. We can then  write the integrand in the region $n<0$ and  $|n|>3(1+w)$ as
\bea\label{BIO}
\left[1+\frac{\kappa^2C_0}{6\ell}a^{-(n+3(1+w))}+\frac{\Lambda}{3\ell}a^{-n}\right]^{-\frac{1}{2}}
=1+\sum_{k=1}^{\infty}\binom{-\f{1}{2}}{k}\left[\frac{\kappa^2C_0}{6\ell}a^{-(n+3(1+w))}+\frac{\Lambda}{3\ell}a^{-n}\right]^k,
\eea
leading to
\bea\label{INTBIO}
\sqrt{\ell}(t_i-t)=-\f{2}{n}\left[a(t)^{-\f{n}{2}}-a_i^{-\f{n}{2}}\right]
+\sum_{k=1}^{\infty}\binom{-\f{1}{2}}{k}\int_{a_i}^{a(t)}da\, a^{-\f{n+2}{2}}{\cal F}(a),
\eea
where
\be
{\cal F}(a)=\left[\frac{\kappa^2C_0}{6\ell}a^{-(n+3(1+w))}+\frac{\Lambda}{3\ell}a^{-n}\right]^k.
\ee
Next we proceed to expand ${\cal F}(a)$ which reads
\bea\label{EXSQB}
{\cal F}(a)=\sum_{j=0}^{k}\binom{k}{j}\left(\f{\kappa^2C_0}{6}\right)^j\f{\Lambda^{k-j}}{3^{k-j}\ell^k}a^{-(nk+3j(1+w))}.
\eea
Substituting (\ref{EXSQB}) into (\ref{INTBIO}) and performing the integration we finally get
\lipsum[0]
\begin{widetext}
\bea\label{FinalSol}
\sqrt{\ell}(t_i-t)&=&\nn-\f{2}{n}\left[a(t)^{-\f{n}{2}}-a_i^{-\f{n}{2}}\right]\!\!\!&-&\!\!\!\sum_{k=1}^{\infty}\sum_{j=0}^{k}\binom{-\f{1}{2}}{k}\binom{k}{j}\!\!\!\left(\f{\kappa^2C_0}{6}\right)^j\!\!\!\f{\Lambda^{k-j}}{3^{k-j}\ell^k}\left\{\f{a(t)^{-\left(n\left(k+\f{1}{2}\right)+3j(1+w)\right)}-a_i^{-\left(n\left(k+\f{1}{2}\right)+3j(1+w)\right)}}{n\left(k+\f{1}{2}\right)+3j(1+w)}\right\}.
\eea
\end{widetext}
\lipsum[0]
Now, if we retain only the terms with $k=1$ in the double summation in (\ref{FinalSol}) we find
\bea\label{KEEP}
\frac{1}{A}\bigg\{\alpha_1\left[a(t)^{-\frac{3}{2}(n+2(1+w))}-a_i^{-\frac{3}{2}(n+2(1+w))}\right]
+\alpha_2\left[a(t)^{-\frac{n}{2}}-a_i^{-\frac{n}{2}}\right]
+\alpha_3\left[a(t)^{-\frac{3n}{2}}-a_i^{-\frac{3n}{2}}\right]\bigg\}
=-(t-t_i),
\eea
where
\bea\label{CEFF}
A=18\ell^{\frac{3}{2}}n(n+2(1+w)),~~\alpha_1=nC_0\kappa^2,
\alpha_2=-36\ell(n+2(1+w)),~~\alpha_3=2\Lambda(n+2(1+w)).
\eea
As the time of the singularity is approached ($t\rightarrow t_s$) the scale factor must vanish, which is guaranteed if
$n<0$ and  $|n|>2(1+w)$. The latter condition, $|n|>3(1+w)$, is sufficient to ensure that $|n|>2(1+w)$. The time at which the singularity forms is therefore\vspace{-2.6mm}
\bea\label{TS}
t_s\approx t_i+\frac{a_i^{-\frac{3}{2}(n+2(1+w))}}{18\ell^{\frac{3}{2}}n(n+2(1+w))}
\left[nC_0\kappa^2-2a_i^{3(1+w)}(n+2(1+w))(18\ell a_i^n-\Lambda)\right].
\eea

\section{Spin effects on the collapse dynamics}\label{DAH}

\subsection{Singular solutions}\label{SS}

We are now in a position to examine whether the singularity occurring in the collapse
setting presented in the previous section is hidden behind a horizon or is visible to external observers. The singularity is covered within an event horizon if trapped surfaces
emerge early enough before the singularity formation and may be visible if the apparent horizon,
which is the outermost boundary of trapped surfaces, fails to form or is delayed
during the collapse process. The key factor that
determines the dynamics of the apparent horizon is the Misner-Sharp
energy \cite{MSE} which describes the mass enclosed within the shell labeled by $r$ at the time $t$, and is defined as \cite{Hay}
\bea\label{MSMASS}
M(t,r)&=&\f{R(t,r)}{2}\left[1+g^{\mu\nu}\partial_{\mu}
R(t,r)\partial_{\nu} R(t,r)\right]\nn
&=&\f{R(t,r)\dot{R}(t,r)^2}{2}.
 \eea
It is worth mentioning that in our  study the effect of the torsion is to add extra spin-dependent terms to the energy momentum tensor, which in turn react on the spacetime geometry. It thus
affects the dynamics of the apparent horizon. The spacetime is
said to be trapped, untrapped and marginally trapped if,
respectively
\be
\f{2M(t,r)}{R(t,r)}>1,~~~~\f{2M(t,r)}{R(t,r)}<1,~~~~\f{2M(t,r)}{R(t,r)}=1.
\ee
The field equations (\ref{FEs00}) and (\ref{FEs11}) can
then be rewritten as \cite{masseqs2}
\bea\label{rewritten}
\f{2M^\prime(t,r)}{R^2R^{\prime}}&=&\rho_{{\rm
eff}}(t)+\rho_{\Lambda}\equiv\rho_{_{{\rm total}}}(t),\nn
-\f{2\dot{M}(t,r)}{R^2\dot{R}}&=&p_{{\rm eff}}(t)+p_{\Lambda}\equiv p_{_{{\rm total}}}(t), \eea
From equations (\ref{MSMASS}) and (\ref{HH}) we readily get

\be\label{2MR}
\f{2M(t,r)}{R(t,r)}=\f{r^2}{3}
\left[3\ell a(t)^{n+2}+\f{\kappa^2}{2}C_0a(t)^{-(1+3w)}+\Lambda
a(t)^2\right].
\ee
So provided that the solution lies in the allowed region, that is $|n|>3(1+w)$,  $-2<n<0$ and $w<-\f{1}{3}$, then if $(2M/R)<1$ initially, or equivalently
\be\label{regularitycon}
\frac{r^2}{3}\left[3\ell a_i^{n+2}+\f{\kappa^2}{2}C_0a_i^{-(1+3w)}+\Lambda
a_i^2\right]<1,
\ee
(as required by regularity \cite{Joshi}), the ratio will remain less than $1$
until the singularity occurs, and a trapped surface will not form.
Otherwise (if $n<-2$ or $w>-1/3$), the ratio $2M/R$ eventually exceeds $1$ as the collapse proceeds, so that
a trapped surface forms before the singularity and therefore covers it.
From the second part of equation
(\ref{rewritten}), the total pressure can be obtained as
\bea\label{PEFF}
p_{_{{\rm total}}}=\f{\kappa^2}{2}C_0wa(t)^{-3(1+w)}-\ell(n+3)a^n-\Lambda.
\eea
The initial data of the collapsing configuration can be
chosen so that the effective pressure is positive at initial epoch, the moment at which the matter distribution begins to collapse at rest, $\dot{a}(t_i)=0$. This can be achieved if
\be\label{ConPosPressure}
\f{\kappa^2}{2}C_0(1+w)>\ell na_i^{n+3(1+w)}>0.
\ee
Having the above condition satisfied, we see that the first and second terms in the right hand side of (\ref{PEFF}) dominate the third one (for $\Lambda<0$) and the total pressure becomes negative for $-1<w<-1/3$ and $-2<n<0$. We require $\ell>0$ so that the singularity time is real. Therefore, we can deduce that at later stages of the collapse the failure of trapped surfaces to form is accompanied by a negative pressure \cite{Goswami}, which is indeed produced due to the fermion condensation. On the other hand, if $n<-2$ and $w>-\f{1}{3}$, which satisfies the condition for trapped surface formation, the pressure can be initially set to be positive and remains positive up to the final stages of the collapse.

Although the pressure is allowed to take on negative values, the collapse process will
be physically reasonable if the weak energy condition is preserved
throughout the collapse. The weak energy condition (WEC) states that the
energy density as measured by any local observer is non-negative.
Thus, along any non-spacelike vector, the following conditions have
to be satisfied
\be\label{WECrhop} \rho_{_{{\rm total}}}\geq0,~~~~~~\rho_{_{{\rm total}}}+p_{_{{\rm total}}}\geq0.
\ee
We thus have
\bea\label{WEC}
&&\rho_{_{{\rm total}}}=\Lambda+\f{\kappa^2}{2}C_0 a(t)^{-3(1+w)}+3\ell a(t)^n\geq0,\nn
&&\rho_{_{{\rm total}}}+p_{_{{\rm total}}}=\f{\kappa^2}{2}C_0(1+w) a(t)^{-3(1+w)}-n\ell a(t)^n\geq0.\nn
\eea
The initial data can be suitably chosen so as to make the second and third terms in the first inequality dominate initially over the first negative term so that the inequality holds initially. Then as these terms diverge at later times the whole inequality remains valid. Therefore it is enough to satisfy
\be\label{weeek}
\Lambda+\f{\kappa^2}{2}C_0 a_i^{-3(1+w)}+3\ell a_i^n\geq0.
\ee
The second inequality is automatically satisfied since $w>-1$, $\ell>0$ and $n<0$. Therefore, the initial set
up which subsequently guarantees the validity of the weak energy condition has to satisfy (\ref{weeek}).
We note that the second inequality in (\ref{WECrhop}) implies the
validity of the null energy condition.

\begin{figure}
\begin{center}
\includegraphics[scale=0.42]{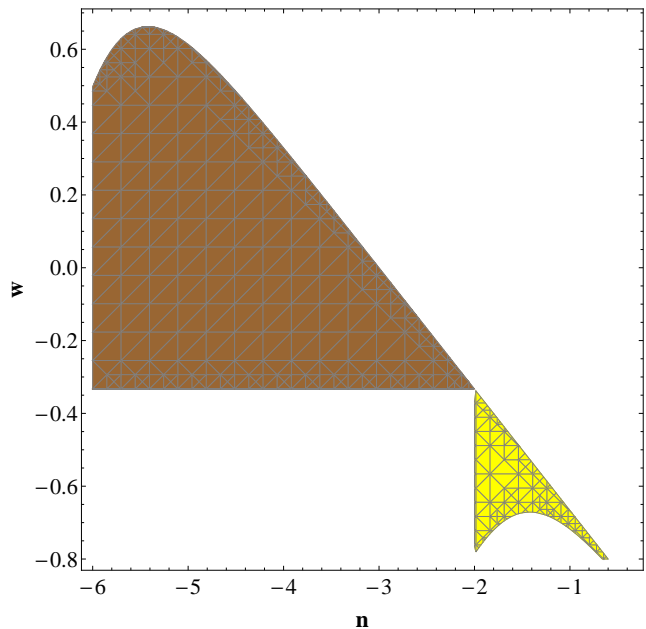}
\caption{The allowed region for $n$ and $w$ for trapped (brown zone) and untrapped regions (yellow zone) for  $\rho_{i_{{\rm SF}}}=0.01$, $a_i=0.1$, $\sigma_0^2=0.00035$ and $J_0^2=0.00275$, $r=0.1$ and $t_i=0$. }\label{Fig1}
\end{center}
\end{figure}

The quantities related to spin source parameters, $\sigma_0^2$ and $J_0^2$, together with those related
to initial values of energy density and scale factor, the barotropic index, $w$ and the exponent of
divergence of spin densities, $n$, all feature in a six-dimensional space of free parameters.
In view of equation (\ref{2MR}), the quantities $n$ and $w$ determine the formation of trapped
surfaces in the dynamical evolution of the collapse. Therefore, any point chosen from the four-dimensional
sub-space ($\sigma_0^2, J_0^2,\rho_{i_{{\rm SF}}},a_i$) constructed by fixing values of $n$ and $w$
represents a collapse scenario that may either end in a black hole or a naked singularity. However,
not all the points in this sub-space are suitable. For the sake of physical reasonableness
the initial configuration should satisfy the regularity condition, (\ref{regularitycon}), the weak energy
condition (\ref{weeek}) and the positivity of total initial pressure
(\ref{ConPosPressure}). That is, the collapse solution starting from the
four-dimensional space mentioned above should respect these conditions. Recall that we demand
$\ell>0$ and we choose the initial data so that $C_0>0$ (the case with $C_0>0$ and $\ell<0$ shall be presented as the bouncing solutions). To ensure the physical reasonability  of the
collapse scenario we require $t_s>t_i$. Figure \ref{Fig1} presents numerically
the two-dimensional sub-space of the allowed region of $n$ and $w$ parameters for fixed values of $\sigma_0^2$, $J_0^2$,  $\rho_{i_{{\rm SF}}}$ and $a_i$. We note that the regions of the two-dimensional parameter space for the formation or otherwise of trapped surfaces, as simply given by (\ref{2MR}), get more restricted due to the physical reasonableness of the collapse setting.

Finally, referring back to  equation (\ref{rhonew}), the only case which we have already excluded \footnote{Through the whole paper, we are excluding this case unless it is explicitly mentioned.} is the case for which $w=0$ and $n=-3$. One can treat this situation by solving  equations (\ref{FEs00}-\ref{CE}) for these values to find
\bea
\!\!\!\!\!\!\!\!\rho_{{\rm SF}}(a)&=&\rho_{i_{{\rm SF}}}\left(\f{a_i}{a}\right)^3+\f{3\kappa^2}{8a^3}\left(\sigma_0^2-48J_0^2\right)\log\left({\f{a}{a_i}}\right),\\
H^2&=&\f{C_2}{a^3}+\f{\kappa^4}{16a^3}\left(\sigma_0^2-48 J_0^2\right)\log(a)+\f{\Lambda }{3}\label{NewH},
\eea
where
\bea
C_2=-\f{\kappa^4}{48}\left(\sigma_0^2+48J_0^2\right)-\f{\kappa^4}{16}\left(\sigma_0^2-48J_0^2\right)\log(a_i)
+\f{\kappa^2}{6}a_i^3\rho_i.
\eea
Looking at the form of  equation (\ref{NewH}), it obviously results in the formation of trapped surfaces as the scale factor approaches zero. Taking all discussions about the physical reasonableness and energy conditions in this section as granted, demanding a physical solution will impose the condition $\sigma_0^2\leq 48J_0^2$ on the spin properties of the matter. In other words, this solution corresponds to a four-dimensional subspace, ($w=0$, $n=-3$), of the six-dimensional parameter space of the problem which is also constrained by the condition $\sigma_0^2\leq 48J_0^2$. More interestingly, for the case where $\sigma_0^2= 48J_0^2$ corresponding to a macroscopic dust fluid, the collapse ends in a black hole with the exterior solution, after a suitable spacetime matching, as that of the Schwarzchild-anti-de Sitter spacetime, see  section \ref{Exterior} for more details.

\subsection{Strength of the naked singularity}\label{Strength}

In order to make  our discussion in the previous
subsection more concrete we need to investigate the curvature strength of the naked
singularity which is an important aspect of its physical nature and
geometrical importance. The main underlying idea is to examine the
rate of curvature growth along non-spacelike geodesics ending at the
singularity,  in the limit of an observer approaching it. The singularity is
said to be gravitationally strong in the sense of Tipler
\cite{Tipler} if every collapsing volume element is crushed to zero
size at the singularity, otherwise it is known as weak. It is widely
believed that when there is a strong curvature singularity forming,
the spacetime cannot be extended through it and is geodesically
incomplete. While if the singularity is gravitationally weak it
may be possible to extend the spacetime through it classically. In
order that the singularity be gravitationally strong there must
exist at least one non-spacelike geodesic with tangent vector
$\xi^{\mu}$, along which the following condition holds in the limit
mentioned above \be \label{SCC}
\Psi=\lim_{\lambda\rightarrow0}\lambda^2R_{\mu\nu}\xi^{\mu}\xi^{\nu}>0,
\ee
where $R_{\mu\nu}$ is the Ricci tensor and $\lambda$ is an
affine parameter which vanishes at the singularity.

Let us now consider a radial null geodesic with tangent vector
$\xi^{\mu}=dx^{\mu}/d\lambda=(\xi^{t},\xi^{r},0,0)$ that terminates at the singularity at $\lambda=0$. We note that
since $\xi^{\mu}$ is  {an affinely parametrized null geodesic}, we
have \be\label{GEO}
\xi^{\mu}\xi_{\mu}=0,~~~~~~~~~~~\xi^{\mu}\nabla_{\mu}\xi^{\nu}=0.
\ee From the null condition for $\xi^{\mu}$, with the help of
the spacetime metric (\ref{metric}), we  obtain \bea\label{SPM}
\f{dt}{dr}=\f{\xi^{t}}{\xi^{r}}=a(t), \eea while the geodesic
equation results in the following differential equations
\bea\label{GEQ}
\xi^{t}\dot{\xi}^{t}+a\dot{a}(\xi^{r})^2=0,\nonumber\\
\xi^{t}\dot{\xi}^{r}+2\f{\dot{a}}{a}\xi^{t}\xi^{r}=0, \eea
which give the vector field tangent to the null geodesics as \be\label{TVF}
\xi^{\mu}=(a^{-1},a^{-2},0,0). \ee Next we proceed to check the
quantity given by (\ref{SCC}) which, with the use of field
equation (\ref{FEs}), reads
\bea\label{SCCRW}
\Psi\propto\lim_{\lambda\rightarrow0}\lambda^2T_{\mu\nu}\xi^{\mu}\xi^{\nu}
=\lim_{\lambda\rightarrow0}\lambda^2\left[T_{tt}(\xi^{t})^2+T_{rr}(\xi^{r})^2\right]
=\lim_{\lambda\rightarrow0}\lambda^2\left[\f{\kappa^2C_0(1+w)}{2a^{(5+3w)}}-\f{n\ell}{a^{2-n}}\right],
\eea
where use has been made of equations (\ref{SPM}) and (\ref{TVF})
and the null energy condition (\ref{WEC}). Next, we proceed by noting that
\be\label{note}
\f{d}{d\lambda}a^{\delta}=\delta Ha^{\delta-1},~~~\f{d^2}{d\lambda^2}a^{\delta}=\delta a^{\delta-2}\left[\dot{H}+(\delta-1)H^2\right].
\ee
Consequently, we find that
\bea\label{Hop}
\Psi\propto\lim_{\lambda\rightarrow0}\Bigg\{\f{\kappa^2C_0(1+w)}{(5+3w)a^{3(1+w)}\left[\dot{H}+
(4+3w)H^2\right]
}
-\f{2n\ell}{(2-n)a^{-n}\left[\dot{H}+(1-n)H^2\right]}\Bigg\}.
\eea
Substituting for the rate of collapse from equation
(\ref{HH}) and noting that the terms $a^{-n}$ and $a^{-(n+3(1+w))}$ go to zero when the scale factor vanishes if $n<0$, $-1<w\leq1$ and $|n|>3(1+w)$, we finally get in the limit of approach to the singularity
\be\label{Psi}
\Psi\propto\f{2|n|}{(2+|n|)\left(1+\f{|n|}{2}\right)}>0.
\ee
Therefore, the strong curvature condition is fulfilled along the singular
null geodesics and the naked singularity is gravitationally strong in the sense of \cite{Tipler}.

\subsection{Non-singular solutions}\label{NSS}
In subsection \ref{SS} we studied the solutions to the field equations that exhibit the formation of spacetime singularity. However, it is expected that the spin effects, which become more important in the very late stages of the collapse procedure, oppose the pull of gravity to balance it.  In such a scenario, the collapse changes to expansion at a turn-around point of the scale factor leading to the singularity removal. Such a class of solutions can be found by setting $\ell<0$ in equation (\ref{HH}). Let us consider the process of collapse that begins at an initial epoch $t_i$ with the initial value of the scale factor $a_i$. As the collapse enters the small scale factor regime at a time, say, $t_{\rm{cr}}>t_i$ with $a_{{\rm cr}}<a_i$, the third term in equation (\ref{HH}) is negligible and we can write (we consider here the dust case with $n=-6$)
\be\label{SSFR}
\dot{a}^2\simeq \ell a(t)^{-4}+\f{\kappa^2C_0}{6}a(t)^{-1},
\ee
for which the solution reads
\bea\label{SB}
a(t)=\Bigg[a_{\rm{cr}}^3+\f{3}{8}C_0\kappa^2(t-t_{\rm{cr}})^2- B(t-t_{{\rm cr}})\Bigg]^{\f{1}{3}},
\eea
where
\be\label{B}
B=\left[9\ell+\f{3}{2}\kappa^2C_0a_{\rm{cr}}^3\right]^{\f{1}{2}}.
\ee
The above solution exhibits a bounce occurring at the finite time $t_{\rm b}$ where the collapse halts ($\dot{a}(t_{{\rm b}})=0$)
at a minimum value of the scale factor given by
\be\label{amin}
a_{{\rm min}}=\left(\f{6|\ell|}{\kappa^2C_0}\right)^{\f{1}{3}}=\left[\f{\kappa^2|\sigma_0^2-48J_0^2|}{8a_i^{3}(\rho_{i_{{\rm SF}}}-12\kappa^2J_0^2a_i^{-6})}\right]^{\f{1}{3}}.
\ee
It is worth noting that the bouncing solutions obtained here stand for $\ell<0$ and $C_0>0$ or equivalently $\sigma_0^2>48J_0^2$ and $\rho_{i_{{\rm SF}}}>12\kappa^2J_0^2a_i^{-6}$, in contrast to the singular ones. 
Next, we proceed with investigating the dynamics of the apparent horizon, that its radius at each instance of time is given by the condition $2M(t,r_{{\rm ah}}(t))=R(t,r_{{\rm ah}}(t))$ or correspondingly
\be\label{SFAH}
r_{{\rm ah}}(t)=\left[\f{\ell}{a(t)^4} +\f{\kappa^2C_0}{6a(t)}+\f{\Lambda}{3}a(t)^2\right]^{-\f{1}{2}}.
\ee
It can now be easily checked that the apparent horizon curve has a minimum for
\be\label{astar}
a_{\star}=\f{1}{2}\left[\f{C_0\kappa^2}{\Lambda}\left(1-\sqrt{1-\f{8\Lambda|\sigma_0^2-48J_0^2|}{C_0^2}}\right)\right]^{\f{1}{3}},
\ee
whence we can find the minimum radius $r_{{\rm min}}=r_{{\rm ah}}(a_{\star})$ so that if the boundary of the collapsing volume is chosen as $r_{\Sigma}<r_{{\rm min}}$, then no horizon would form during the collapsing and expanding regimes. Correspondingly, from the first part of equation (\ref{rewritten}) we can define a threshold mass,  setting $m(t,r)=2M(t,r)$
\be\label{THRM}
m_{\star}=m(a_{\star},r_{{\rm min}})=\f{r_{{\rm min}}^3}{3}\left[3\ell a_{\star}^{-3}+\f{\kappa^2}{2}C_0+\Lambda a_{\star}^3\right],
\ee
in such a way that if $m<m_{\star}$ then the formation of apparent horizon is avoided.

\begin{figure}
\begin{center}
\includegraphics[scale=0.58]{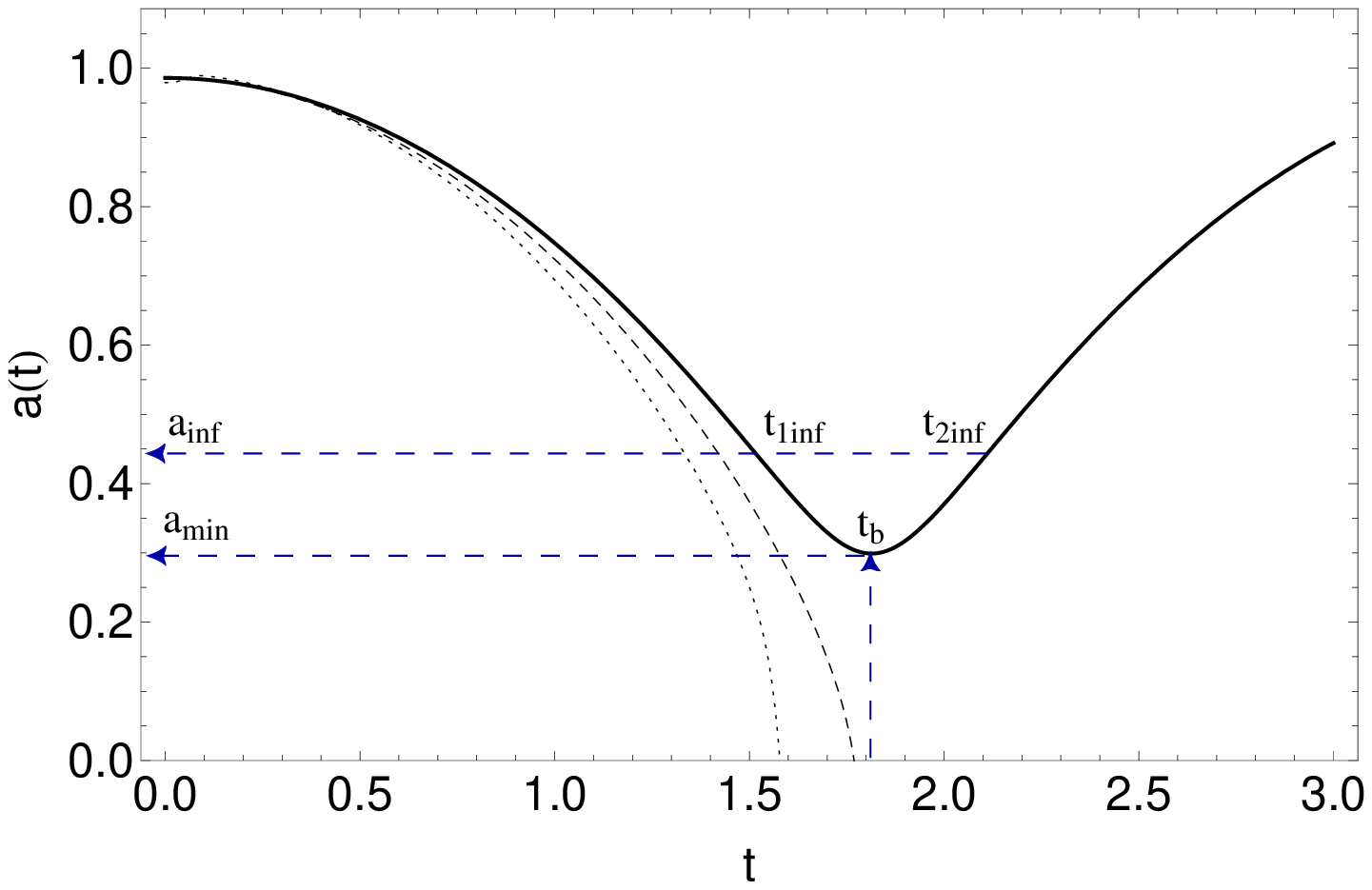}
\includegraphics[scale=0.5]{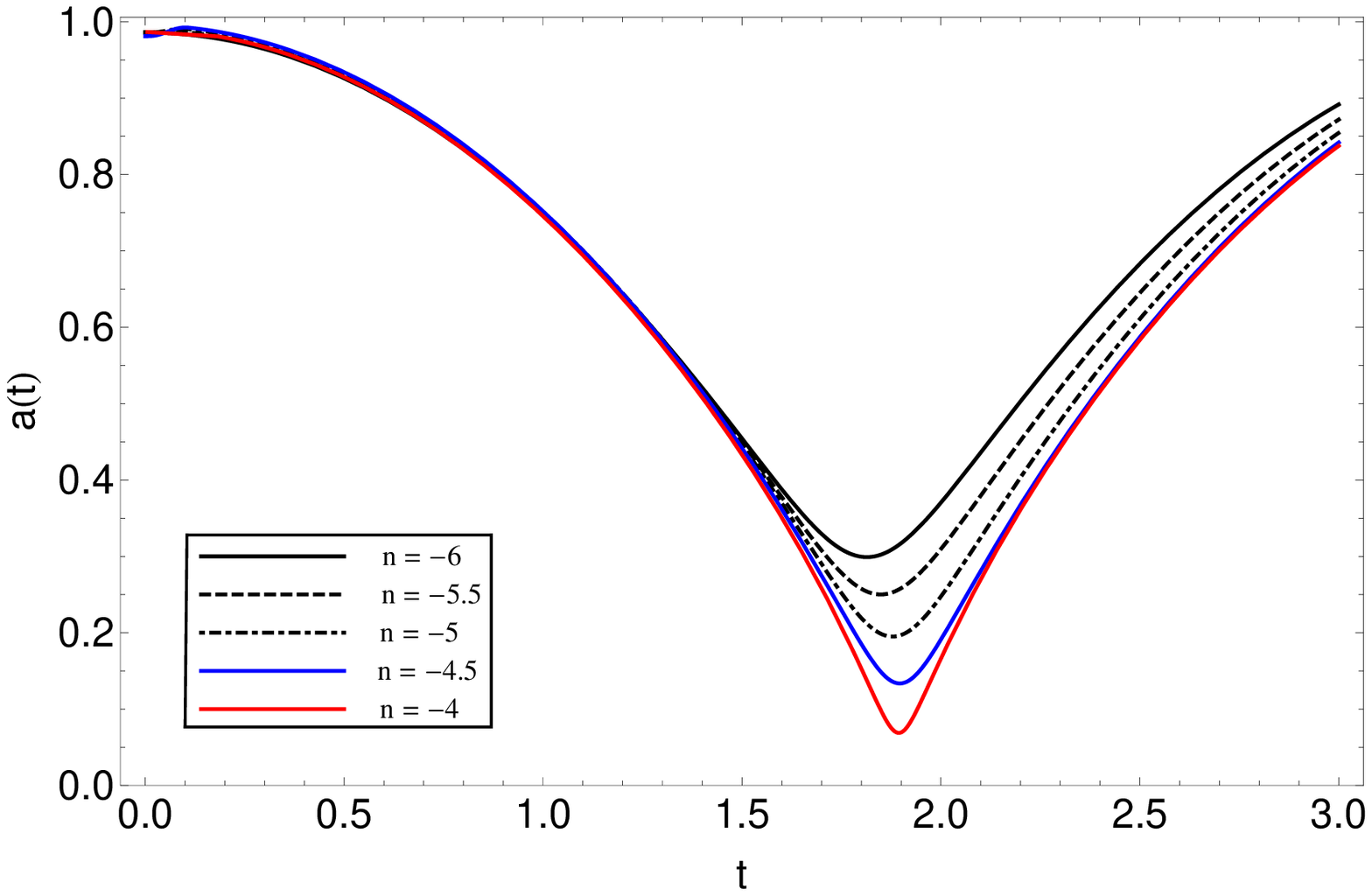}
\caption{Left panel: Time behavior of the scale factor for $a_i=0.986$, $n=-6$, $w=0$ and $\sigma_0^2=0.89$, $J_0^2=0.01$, $\rho_{i_{{\rm SF}}}=2.034$ (solid curve), $\sigma_0^2=J_0^2=0$, $\rho_{i_{{\rm SF}}}=2$ (dashed curve) and $\sigma_0^2=0.35$ and $J_0^2=0.011$ (dotted curve). The first and second inflection points occur at the times $t_{{\rm 1inf}}=1.513$ and $t_{{\rm 1inf}}=2.113$, respectively. The scale factor at this point takes the value $a_{{\rm inf}}=0.445$. The bounce occurs at the time $t_{{\rm b}}=1.813$ for which the scale factor takes the value $a_{{\rm min}}=0.299$. Right panel: Time behavior of the scale factor for different values of spin density divergence parameter.}\label{sftb}
\end{center}
\end{figure}

In figure \ref{sftb}, we present numerically the trajectory of the scale factor featuring the occurrence of a bounce, for
different values of the space parameters. As it is shown in the left panel of figure \ref{sftb} for $\ell=0$, the collapse progresses till the singularity formation (dashed curve) while for $\ell<0$ (solid curve) the collapsing matter bounces back at a finite value of the scale factor.
The dotted curve represents a case, with $\ell>0$ case in which the singularity happens sooner than the case where the
effects are totally excluded ($\ell=0$). The right panel emphasizes the role of spin contributions in the time behavior of the scale
factor. For larger values of $n$, it takes longer for spin effects to become
strong enough to prevent the collapse, which consequently happens at smaller radii.
It is also seen, from equation (\ref{amin}) that for a fixed value of  $n$ the larger the initial energy
density the smaller the minimum value of the scale factor at which the bounce occurs.

\begin{figure}
\begin{center}
\includegraphics[scale=0.55]{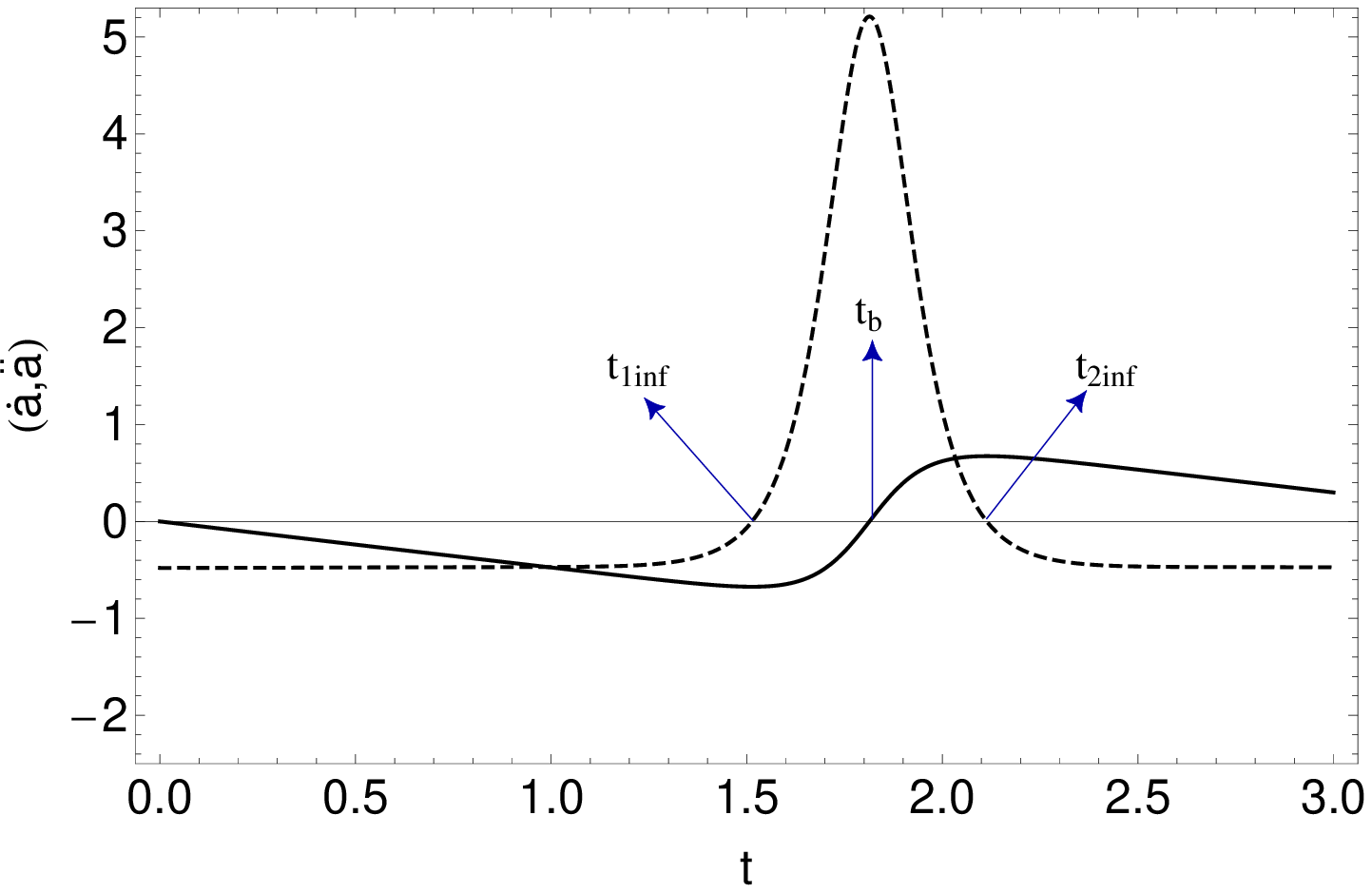}
\includegraphics[scale=0.62]{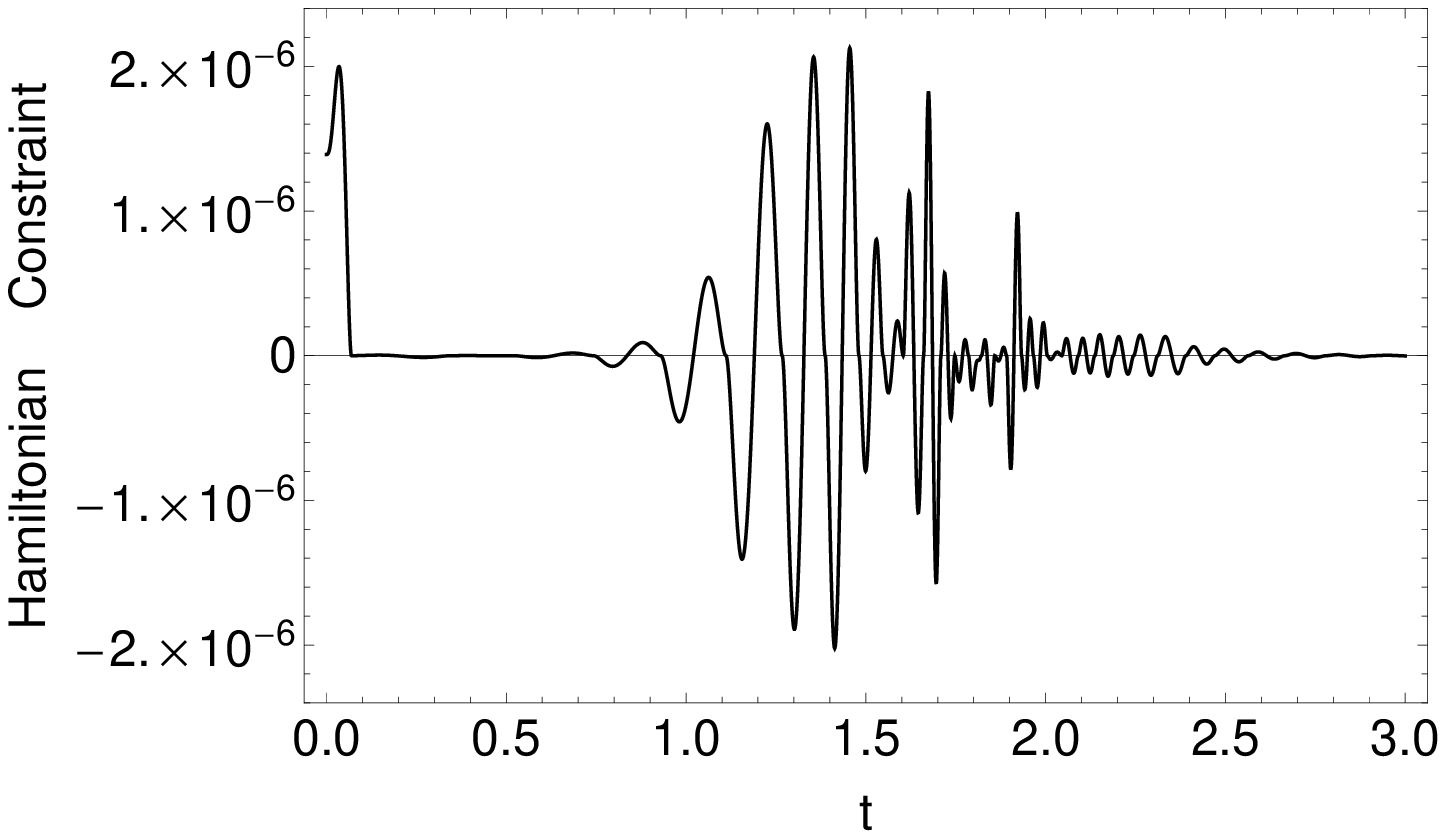}
\caption{Left panel: Time behavior of the collapse velocity (solid curve) and acceleration (dashed curve) for $a_i=0.986$, $\rho_{i_{{\rm SF}}}=2.034$, $n=-6$, $w=0$, $\sigma_0^2=0.89$ and $J_0^2=0.01$. Right panel: The Hamiltonian constraint during the whole evolution of the system.}\label{adadd}
\end{center}
\end{figure}

The left panel in figure \ref{adadd} further illustrates that the dynamical evolution can be divided to four regimes, two of which during the contracting and the rest during the expanding phases. The matter volume begins to collapse from rest ($\dot{a}(t_i)=0$ and $\ddot{a}(t_i)<0$), immediately entering an accelerated contracting phase, called \lq\lq{}{\it fast reacting}\rq\rq{} regime, until the small scale factor regime is reached where a decelerated contracting phase starts as a result of spin domination, that is a \lq\lq{}{\it slow
reacting}\rq\rq{} regime. The point at which transition between these two phases occurs corresponds to the first inflection
point where $\ddot{a}(t_{{\rm 1inf}})=0$ and the collapse velocity reaches its maximum value
 $|\dot{a}|_{{\rm max}}=|\dot{a}(t_{{\rm inf}})|$. The collapse then ceases to proceed at the bounce where
$\dot{a}(t_{{\rm b}})=0$ and the scale factor approaches its minimum value $a_{{\rm min}}=a(t_{\rm b})$. At this moment, the acceleration reaches its absolute maximum value $\ddot{a}(t_{\rm b})>0$. The early stages of the
post-bounce evolution are controlled by an inflationary expanding phase (or a\lq\lq{}{\it fast reacting}\rq\rq{} regime (of expansion)) until the small scale factor regime ends where the acceleration curve reaches its second inflection point
($\ddot{a}(t_{{\rm 2 inf}})=0$) and the velocity reaches its maximum with the same absolute value as in the contracting phase (but actually positive). Afterward a decelerated expanding phase governs the scenario (an expansionary \lq\lq{}{\it slow reacting}\rq\rq{} regime).

In the right panel of figure \ref{adadd}, we plot the Hamiltonian constraint (\ref{HH}) throughout the
dynamical evolution of the collapsing object as governed by equation (\ref{addot}). We see that this constraint is numerically satisfied with
the accuracy of the order of $10^{-6}$ or less. 

\begin{figure}
\begin{center}
\includegraphics[scale=0.47]{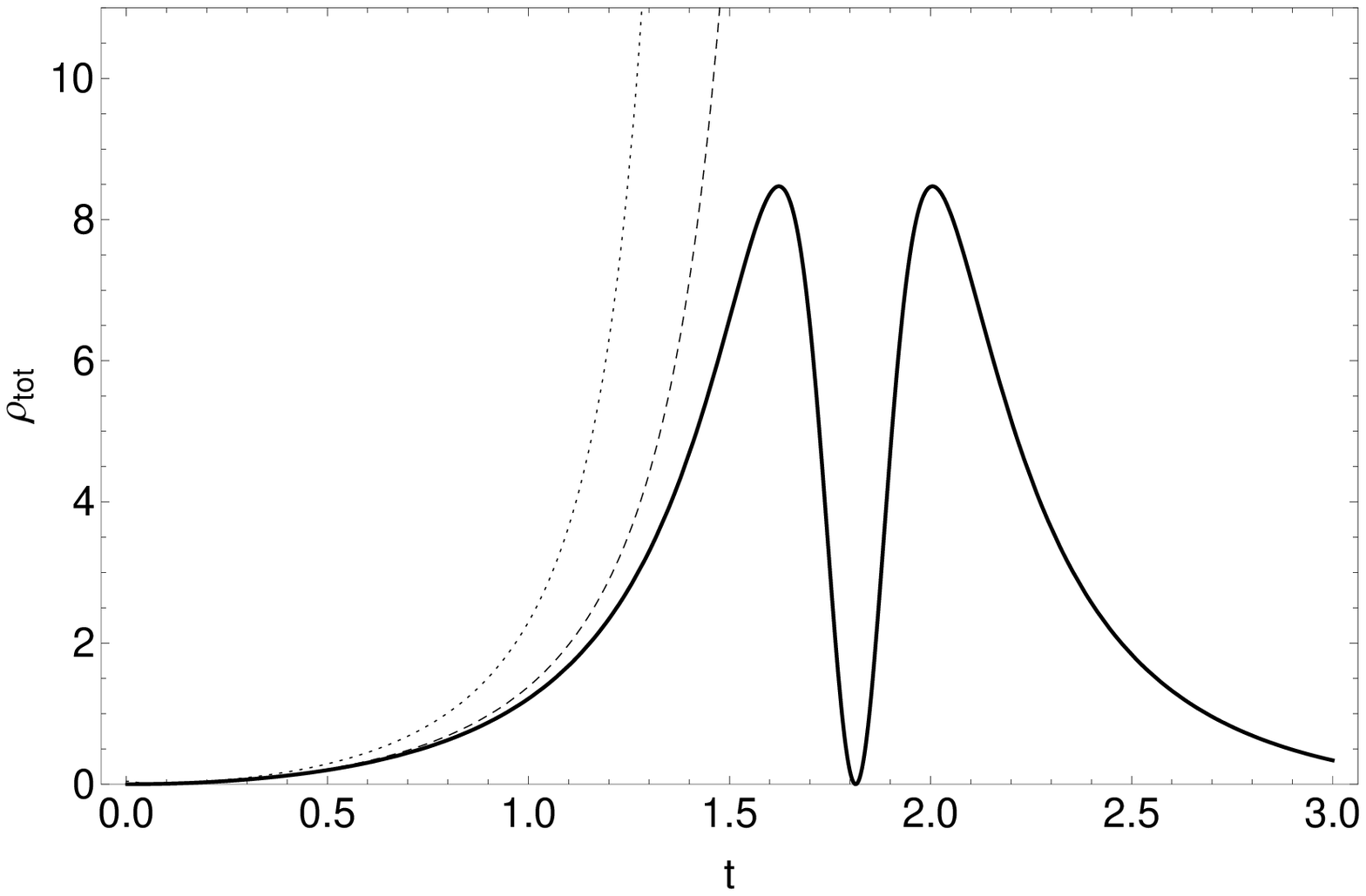}
\includegraphics[scale=0.48]{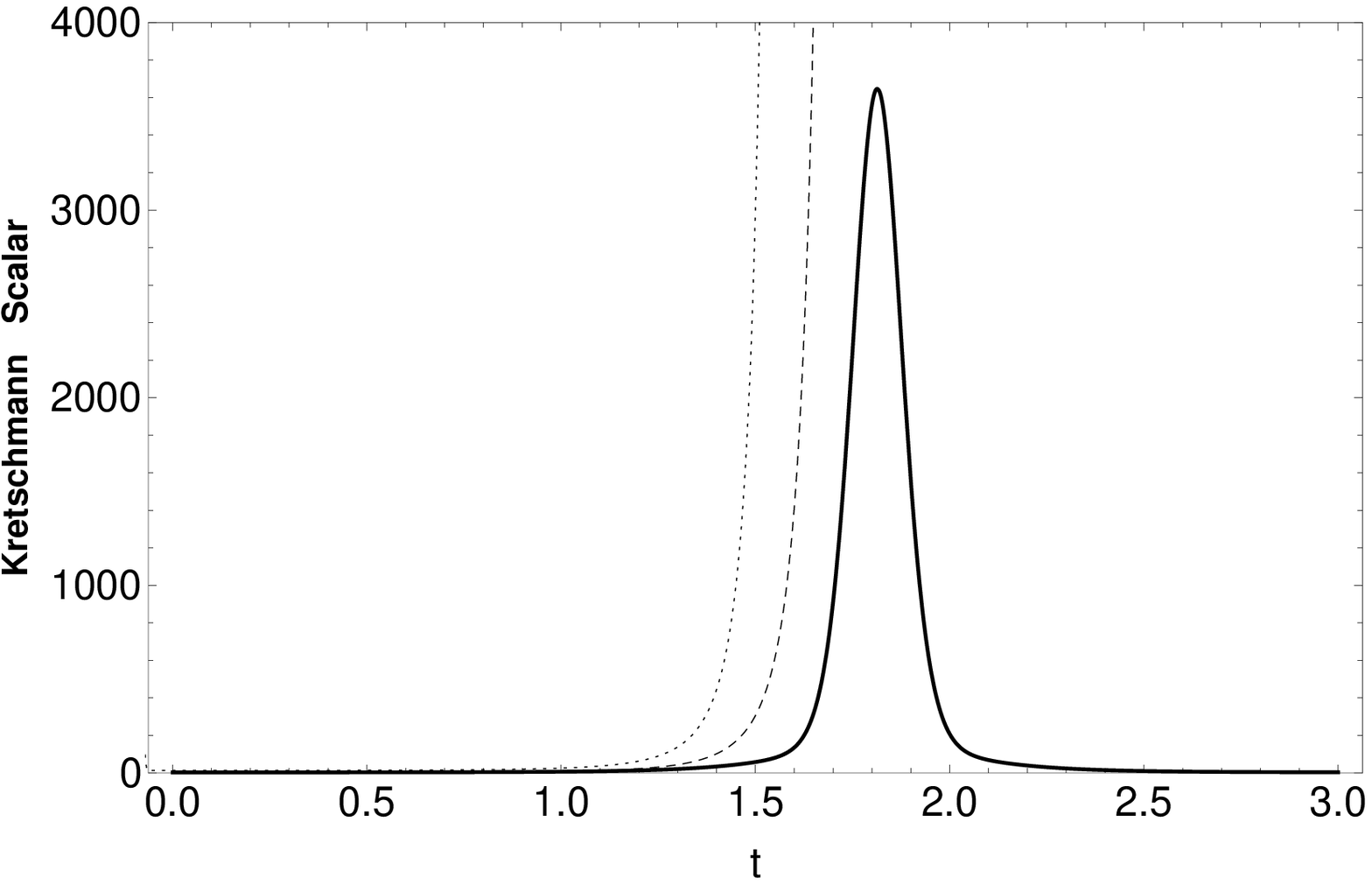}
\caption{Left panel: Time behavior of the total energy density for $a_i=0.986$, $\rho_{i_{{\rm SF}}}=2.034$, $n=-6$, $w=0$ and $\sigma_0^2=0.89$, $J_0^2=0.01$ (solid curve), $\sigma_0^2=J_0^2=0$ (dashed curve) and $\sigma_0^2=0.35$ and $J_0^2=0.011$ (dotted curve). Right panel: Time behavior of the Kretschmann scalar for the same parameters as above.}\label{rhowec}
\end{center}
\end{figure}

The left panel in figure \ref{rhowec} shows the behavior of total energy
density as a function of time. At the beginning of the collapse, the spin contribution is insignificant, while as the collapse
advances, the gravitational attraction succumbs to the spin density correction term which comes into play and behaves as a negative energy density. 
The solid curve shows that the total energy density increases up to its first maximum, then it
decreases suddenly, owing to the dominating negative energy density coming from spin correction term, and tends to zero at
the bounce time at which $\dot{\rho}_{{\rm tot}}(t_{{\rm b}})=0$. This tells us that the matter content within the collapsing
volume becomes incompressible. At the post-bounce regime the spin density term becomes diluted as a consequence of the
inflationary expanding phase causing the total energy density to increase up to its second maximum and then falls off to finite values. Thus, the total energy density is finite for $\ell<0$ during the dynamical evolution of the collapse scenario while the dashed
($\ell=0$) and dotted ($\ell>0$) curves signal the occurrence of a spacetime singularity where the  Kretschmann scalar (right panel) and energy density diverge.

\begin{figure}
\begin{center}
\includegraphics[scale=0.57]{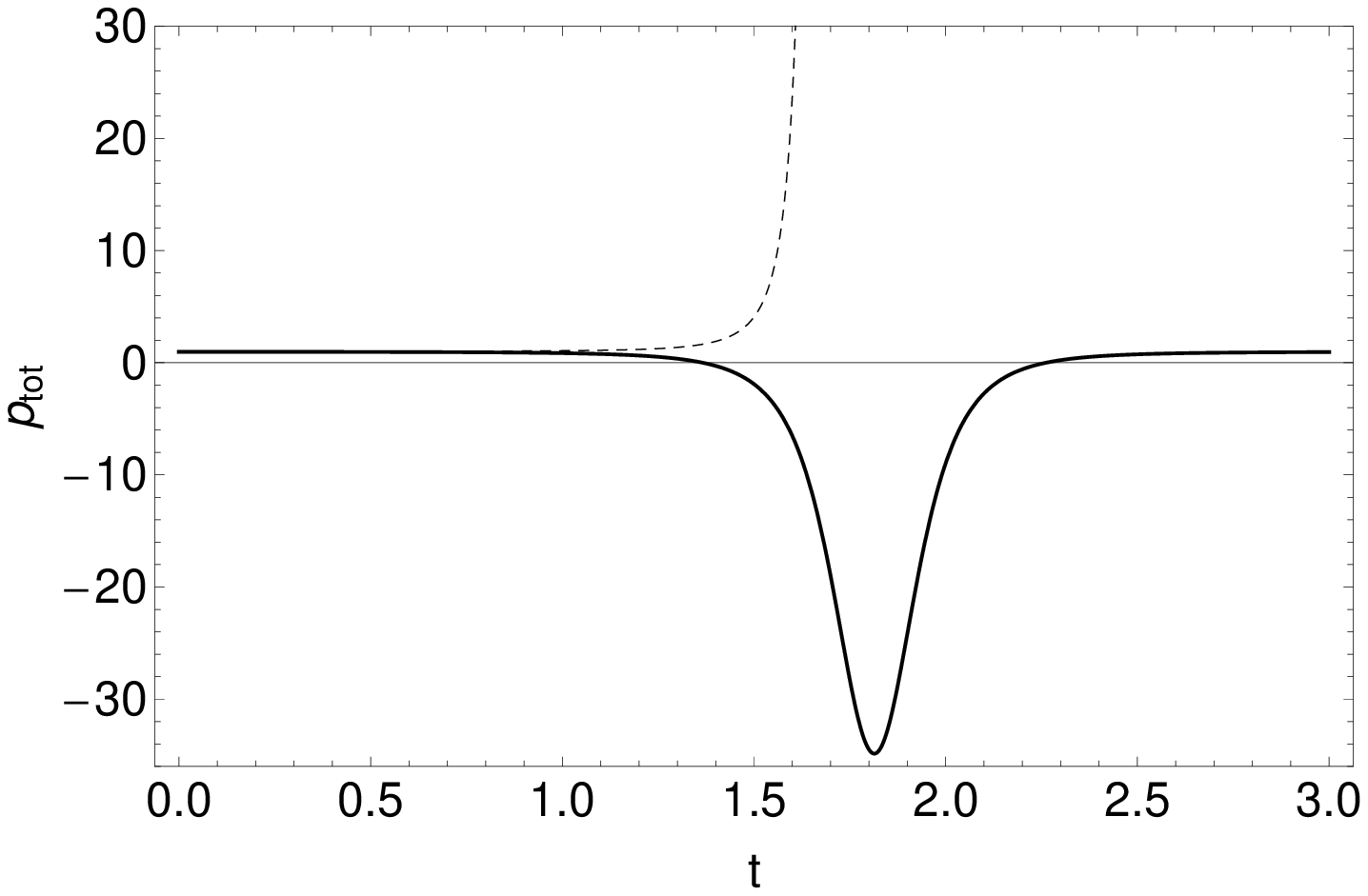}
\includegraphics[scale=0.56]{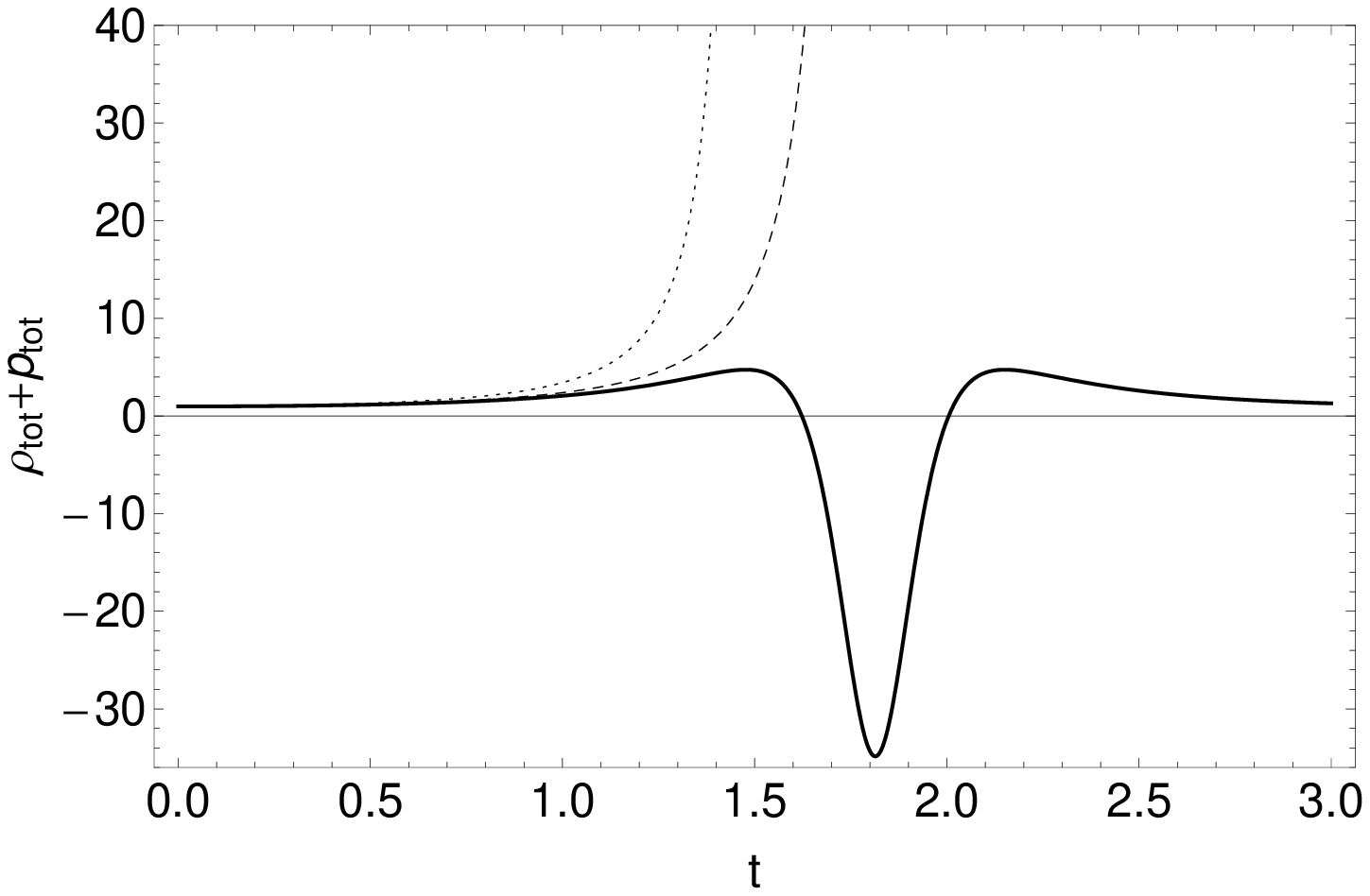}
\caption{Left panel: Time behavior of the total pressure for $a_i=0.986$, $\rho_{i_{{\rm SF}}}=2.034$, $n=-6$, $w=0$ and $\sigma_0^2=0.89$, $J_0^2=0.01$ (solid curve), $\sigma_0^2=0.35$ and $J_0^2=0.011$ (dashed curve). Right panel: The weak energy condition for the same parameters as above.}\label{pwec}
\end{center}
\end{figure}

The left panel in figure \ref{pwec} shows the behavior of total pressure during the
whole dynamical evolution of the collapse. The solid curve ($\ell<0$) shows that the total pressure is positive in the early
stages of the collapse where the spin contribution is weak. As the collapse proceeds the pressure becomes negative and
reaches a maximum value in negative direction, where the contracting phase turns to an expanding one. It is the appearance
of such a negative pressure, as produced by a spin correction term, which causes the bounce. Whereas the dashed curve
($\ell>0$) shows that the pressure begins from a positive value and remains positive up to the singularity formation. The right panel shows that the weak energy condition is satisfied in the absence of spin effects (dashed curve) and also for the case
in which $\ell>0$. For $\ell<0$, WEC holds in the weak field regime while it is violated in the spin dominated regime. Such
a violation of WEC can be compared to the models in which the effects of quantum gravity are taken into account
\cite{WEAKVIO}. 

We also need to check if a dynamical horizon is formed during the whole contracting and expanding phases.
Firstly, as we stated in subsection \ref{SS}, the regularity condition has to be respected at the time at which the collapse
commences. Let us define the maximum radius $r_{\Sigma_{\rm max}}$ in such a way that if
$r_{\Sigma}=r_{\Sigma_{\rm max}}$, then the regularity condition would be violated. Therefore, if the boundary is chosen so
that $r_{{\rm min}}<r_{\Sigma}<r_{\Sigma_{\rm max}}$, four horizons can form.

\begin{figure}
\begin{center}
\includegraphics[scale=0.53]{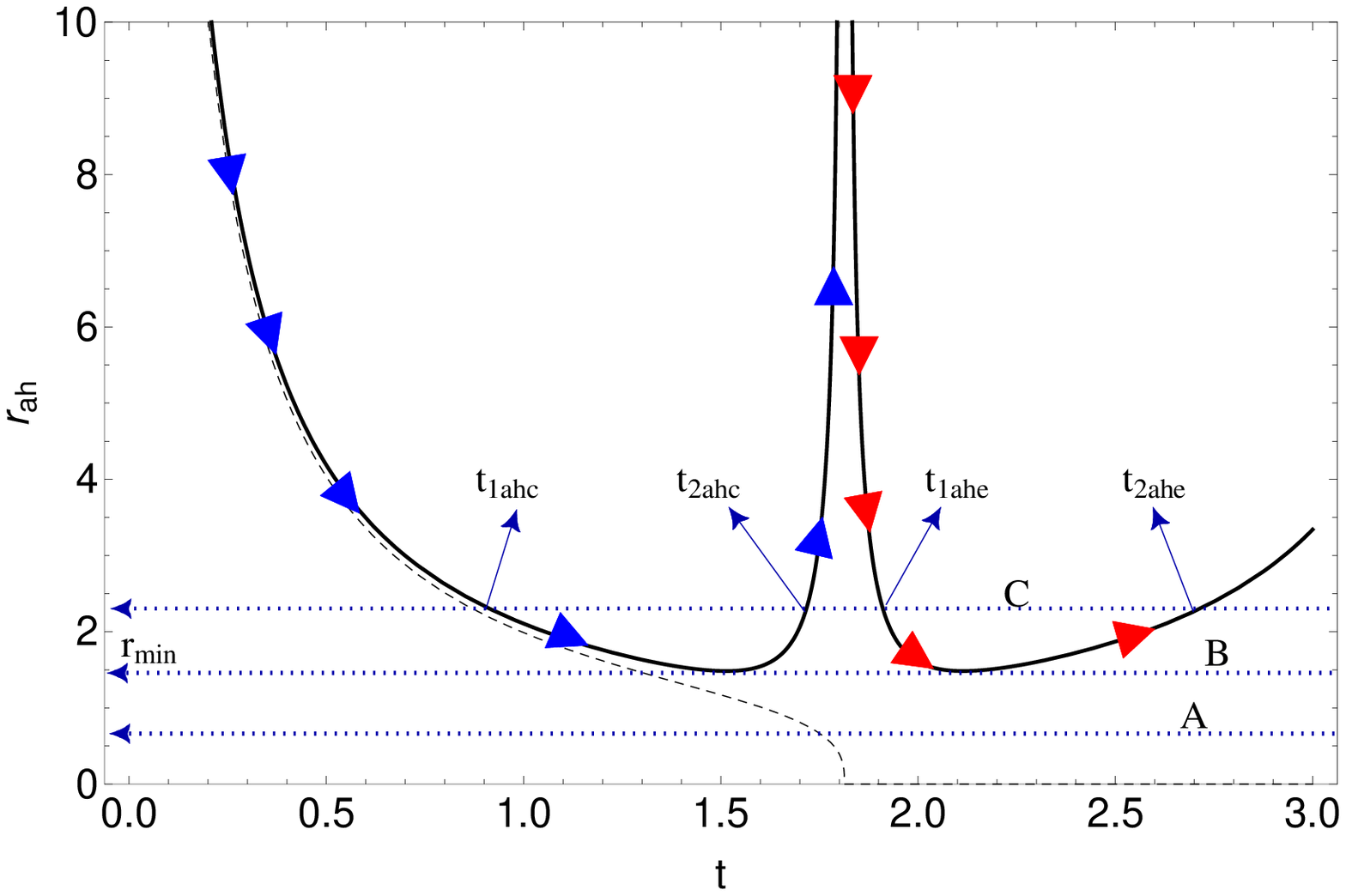}
\includegraphics[scale=0.57]{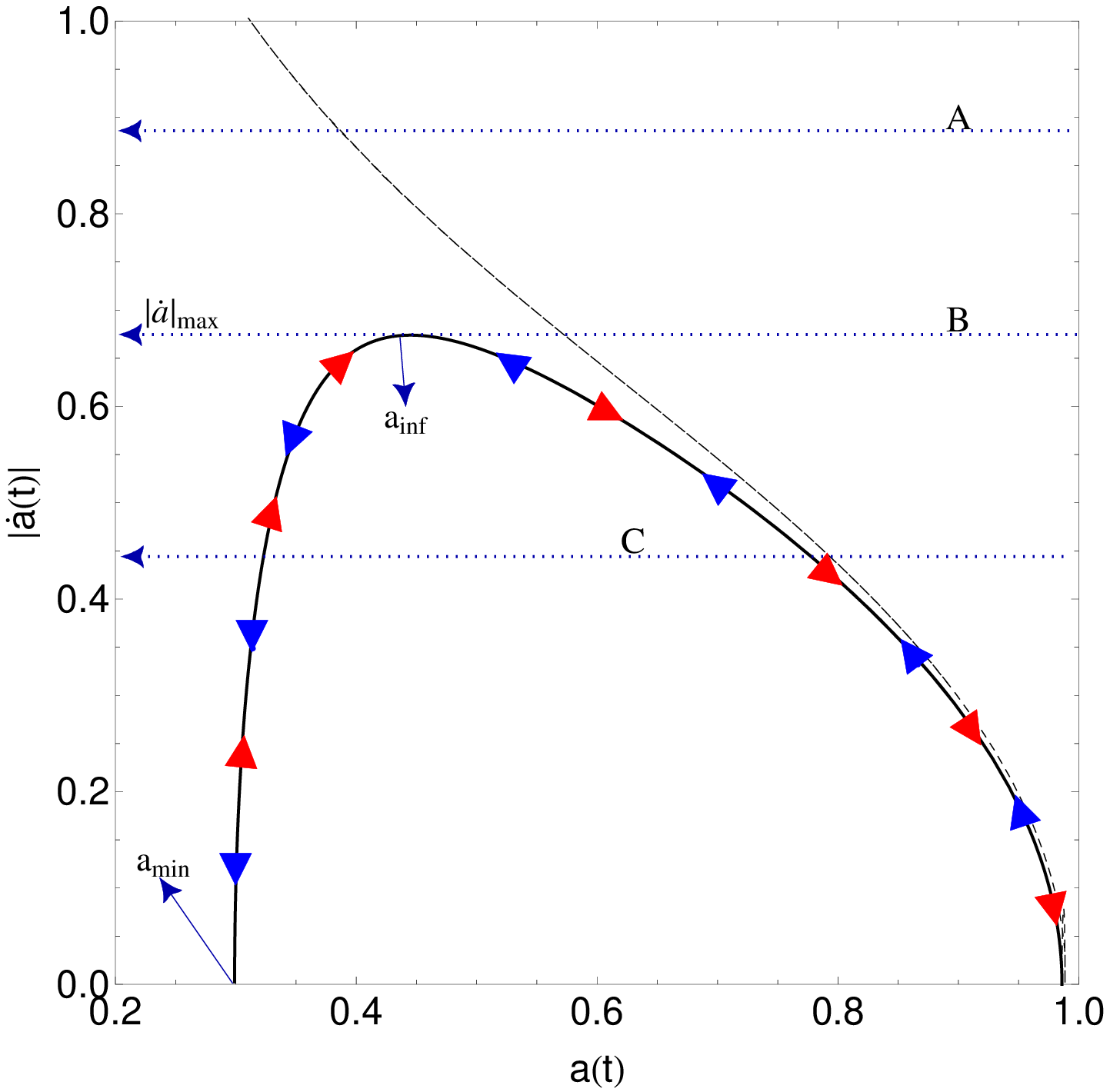}
\caption{Left panel: The behavior of the apparent horizon curve for $a_i=0.986$, $\rho_{i_{{\rm SF}}}=2.034$, $n=-6$, $w=0$ and $\sigma_0^2=0.89$, $J_0^2=0.01$ (solid curve) and $\sigma_0^2=J_0^2=0$, $\rho_{i_{{\rm SF}}}=2$ (dashed curve). The horizontal dotted lines show the times at which four horizons form (C) two horizons form (B) and no horizon forms (A). The times at which the horizons form at the four phases of evolution are
$t_{{\rm 1ahc}}\approx 0.904$, $t_{{\rm 2ahc}}\approx 1.713$, $t_{{\rm 1ahe}}\approx 1.919$ and $t_{{\rm 2ahe}}\approx 2.698$, respectively. There can be found a minimum value for the size of the collapsing object so that the formation of the apparent horizon is prevented (solid curve). This situation cannot be happened when the apparent horizon radius vanishes at a finite amount of time (dashed curve). The blue arrows lying on the solid curve display the direction of collapse process and the red arrows show the direction of expansion process. Right panel: The absolute value of the collapse velocity against the scale factor for the same parameters as above. The horizontal dotted lines intersecting the solid curve show that in case in which the collapse velocity is bounded, four horizons (C), two horizons (B) and no horizon (A) can be formed during the dynamical evolution of the collapse process. The absolute maximum value of the collapse velocity at the inflection points has the value $|\dot{a}|_{{\rm max}}\approx 0.673$. The blue arrows lying on the solid curve show the direction of collapse regime and the red ones show the direction of expanding regime. The formation of an apparent horizon is always guaranteed as the dashed curve shows.}\label{abra}
\end{center}
\end{figure}

The left panel in figure \ref{abra}
shows the behavior of apparent horizon curve (\ref{SFAH}) as compared to the case in which spin effects are absent.  As the
solid curve shows ($\ell<0$), the apparent horizon curve decreases for a while in the contracting phase, and increases just
before the bounce occurs. The first horizon appears in the accelerated contracting phase and, after passing through the
first inflection point, the second one forms in the decelerated contracting phase. As the collapse process turns to an expanding
regime, the apparent horizon decreases again, in the accelerated expanding phase, to meet the boundary for the third time but
at the same radius. After the second inflection point is reached, the scenario is entering a decelerated expanding
phase where the fourth horizon intersects the boundary (see the horizontal dotted line labeled as C). The next possibility for the
horizon formation happens if we take $r_{\Sigma}=r_{{\rm min}}$.
In this case two horizons could form. The first one appears
at the moment of transition between accelerated and decelerated contracting regimes, i.e., the first inflection point. The
second one appears at the same radius but at the time at which the accelerated expanding regime transits to the decelerated
expanding one, i.e., the second inflection point (see the horizontal dotted line labeled as B). Finally, if we take
$r_{\Sigma}<r_{{\rm min}}$ no horizon would form indicating the existence of a minimum value for the size of the
collapsing object so that the formation of the apparent horizon is prevented (see the horizontal dotted line labeled as A).
However, the apparent horizon in the absence of spin effects (dashed curve) propagates inward to finally cover the singularity.
There can be found no minimum for the surface boundary of the collapsing matter in order to avoid the formation of the apparent horizon
and the collapse scenario inevitably results in black hole formation. 

The existence of a minimum value for the size of the collapsing object can
be translated as saying that the speed of collapse has to be limited.  As the solid curve in the right panel in figure \ref{abra} shows, in the
early stages of the collapse, the trajectory of the system in $(|\dot{a}|,a)$ plane follows the dashed curve in which the
 spin effects are neglected. At later times, it deviates from this curve to reach the maximum value for the speed of collapse, i.e., at the first inflection point. After this time, the collapse progresses with a decreasing speed reaching the minimum value of the scale factor, after which the collapsing cloud turns into an expansion. The absolute value of the collapse velocity is bounded throughout the contracting and expanding phases. In this sense, there can be found a maximum value for the collapse velocity (as related to a minimum value for the surface boundary or a threshold mass) so that for $|\dot{a}|>|\dot{a}|_{{\rm max}}$ the horizon equation is never satisfied (see the horizontal dotted line labeled as A). However, if $|\dot{a}|=|\dot{a}|_{{\rm max}}$, two horizons could still appear, first one at the contracting and the second one at the expanding regimes. Both of these horizons form at the same value of the scale factor at inflection points (see the horizontal dotted line labeled as B). The third possibility is $|\dot{a}|<|\dot{a}|_{{\rm max}}$, for which four horizons could appear at the four phases of dynamical evolution of the scenario (see the horizontal dotted line labeled as C). On the other hand, in contrast to these cases, the collapse velocity diverges when the spin effects are absent and the horizon equation is always satisfied, the dashed curve.

Finally, regarding the case $w=-1$, equation (\ref{CE}) leads to the following expression for the energy density
\be
\rho_{{\rm SF}}=\rho_{i_{{\rm SF}}}+\f{\alpha}{n}(a^n-a_i^n),
\ee
whereby from equations (\ref{FEs00}) and (\ref{FEs11}) we can solve for
the collapse rate as
\be
H=\pm\left[H_{i}^2+\kappa^4\f{\sigma_{0}^2}{8n}(a^{n}-a_{i}^{n})\right]^{\f{1}{2}}.
\ee
From the above equation we see that the collapse begins with a contraction phase
(choosing the minus sign) which proceeds until the scale factor reaches
a critical value, $a_c$,
\be
a_{c}=\left[a_{i}^{n}-\f{8n}{\kappa^4\sigma_0^2}H_{i}^2\right]^{\f{1}{n}},
\ee
for which $H(a_c)=0$.
This occurs at the time
\be
t_{c}=t_i-\f{2}{n\sqrt{H_i^2-\f{\kappa^4}{8n} \sigma_0^2 a_i^n}}\arctanh\left[\sqrt{\f{H_i^2}{H_i^2-\f{\kappa^4}{8n}\sigma_0^2 a_i^n}}\right],
\ee
and the Kretchmann invariant and the energy density behave regularly, thus
no singularity forms\footnote{It should be noted that big-rip, sudden or
even type III singularities do not happen here since $\rho$ and $p$ are finite at
$t=t_c$. A type IV singularity does not occur either since the higher
derivatives of $H$ do not diverge at $t=t_c$ \cite{futuresin}.}. To our knowledge this unprecedented situation is a specific feature of the present model. It can be interpreted as a stationary state. Also there is a chance that this unorthodox behavior of the fluid will be followed by a transition to an expansion phase
corresponding to the situation\footnote{It should be noticed
that if no bounce occurs and the collapse goes beyond $a_c$, the effective energy density of the collapsing
object would be negative and, as a result, the weak energy condition
will be violated. The gravitational collapse
of regions with negative energy density has been discussed in the
literature, mainly in the context of topological black holes
\cite{Mann}. It has been claimed that topology changing
processes, due to quantum fluctuations of spacetime, would be a
possible mechanism for such behavior.
However this discussion is beyond the scope of this paper.} where $H$ jumps from a $-$ to a $+$ branch \cite{Malafarina}.

\vspace{0.5cm}
\section{Exterior solution}\label{Exterior}
The gravitational collapse setting studied so far deals with the interior of the collapsing object. We found two classes of solutions, where for the singular ones, depending on the spin source parameters and initial energy density the apparent horizon can be avoided. However, the absence of apparent horizon in the dynamical process of collapse does not necessarily imply that the singularity is naked \cite{WALDAH}. In fact, the singularity is naked if there exist future pointing null geodesics terminating in the past at the singularity. These geodesics have to satisfy ${dt}/{dr}=a(t)$ in the interior space-time so that the area radius must increase along these geodesics. As discussed in \cite{G1}, this situation cannot happen since the singularity occurs at the same time for all collapsing shells. However, this process, to be completely discussed, may require a suitable matching to an exterior region whose boundary $r=r_{\Sigma}$ is the surface of the collapsing matter that becomes singular at $t=t_s$, into which null geodesics can escape. Employing the junction conditions \cite{JCS}, our aim here is to complete the full spacetime geometry presented for the spherically symmetric gravitational collapse via matching the homogeneous interior spacetime
to a suitable exterior spacetime.

Let us therefore consider a timelike three-dimensional hypersurface $\Sigma$ resulting from the isometric pasting of two hypersurfaces $\Sigma^{+}$ and $\Sigma^{-}$, which, respectively, bound the four-dimensional exterior (${\mathcal V}^{+}$) and interior (${\mathcal V}^{-}$) spacetimes. For the interior region we take the line element (\ref{metric}) in the FRW form as
\be \label{int}
ds_{-}^2=dt^2-a^2(t)dr^2-a^2(t)(r^2d\theta^2+r^2\sin^2\theta d\phi^2),
\ee
where the interior coordinates are labeled as $\left\{X_{-}^{\mu}\right\}\equiv\left\{t,r,\theta,\phi\right\}$.
The line element for the exterior region in retarded (exploding) null coordinates is taken as
\be \label{out}
ds_{+}^2=f(v,{\mathcal R})dv^2+2dvd{\mathcal R}-{\mathcal R}^2(d\theta^2+\sin^2\theta d\phi^2),
\ee
where $f(v,{\mathcal R})=1-2{\mathcal M(\mathcal R, v)}/{\mathcal R}$ with ${\mathcal M(\mathcal R, v)}$ being the exterior mass function and the exterior coordinates are labeled as $\left\{X_{+}^{\mu}\right\}\equiv\left\{v,{\mathcal R},\theta,\phi\right\}$. We assume that $\Sigma$ is endowed with an intrinsic line element given by
\be\label{intrinsicmetric}
ds_{\Sigma}^2=d\tau^2-Y^2(\tau)(d\theta^2+\sin^2\theta d\phi^2).
\ee
Here $y^a=\left\{\tau,\theta,\phi\right\}$, $(a=0,2,3)$ are the coordinates of $\Sigma$ with $\tau$ being the time coordinate defined on it and we have chosen the angular coordinates $\theta$ and $\phi$ to be continuous. The governing equations of hypersurface $\Sigma$ in the coordinates $X_{\pm}^{\mu}$ are given by
\be\label{eqsigma}
r-r_{\Sigma}=0~~~~~\text{in}~{\mathcal V}^+,~~~~~~{\mathcal R}-{\mathcal R}_{\Sigma}=0~~~~~ \text{in}~ {\mathcal V}^-.
\ee
Using the above two equations, the interior and exterior induced metrics on $\Sigma^{-}$ and $\Sigma^{+}$ respectively take the form
\be\label{INMIN}
ds_{\Sigma^{-}}^2=dt^2-a^2(t)r_{\Sigma}^2(d\theta^2+\sin^2\theta d\phi^2),
\ee
and
\bea \label{INMEX}
ds_{\Sigma^{+}}^2=\left[f\big(v(t),{\mathcal R}(t)\big)\dot{v}^2+2\dot{{\mathcal R}}\dot{v}\right]dt^2
-{\mathcal R}^2(t)(d\theta^2+\sin^2\theta d\phi^2).
\eea
We assume that there is no surface stress-energy or surface tension at the boundary (see e.g. \cite{SULAYER} for the study of junction conditions for boundary surfaces and surface layers). Then the junction conditions require
\be\label{FFF}
ds_{\Sigma}^2=ds_{\Sigma^{-}}^2=ds_{\Sigma^{+}}^2,
\ee
which gives
\be\label{FFF1}
f\big(v(t),{\mathcal R}(t)\big)\dot{v}^2+2\dot{{\mathcal R}}\dot{v}=1,~~~~~~{\mathcal R}(t)=r_{\Sigma}a(t)=Y(\tau),
\ee
where an overdot denotes $d/dt$. Next, we need to compute the components of the extrinsic curvature of the interior and exterior hypersurfaces. The unit spacelike normal vector fields to these hypersurfaces are given by
\bea\label{NVF}
n^{-}_{\mu}=\left[0,a(t),0,0\right],~~~ n^{+}_{\mu}=\frac{1}{\left[f(v,{\mathcal R})\dot{v}^2+2\dot{{\mathcal R}}\dot{v}\right]^{\frac{1}{2}}}\left[-\dot{\mathcal R},\dot{v},0,0\right].
\eea
Let us take $X^{\mu}=X^{\mu}(y^a)$ as the parametric equations of $\Sigma$. The extrinsic curvature or second fundamental form of the hypersurface $\Sigma$ is a three-tensor defined as \cite{EPRT}
\be\label{EXCDEFINE}
K_{ab}\equiv e_{a}^{\nu}e_{b}^{\mu}\nabla_{\mu}n_{\nu},
\ee
where $e_{a}^{\nu}=\partial X^{\nu}/\partial y^a$ are the basis vectors tangent to the hypersurafce $\Sigma$ and the covariant derivative is taken with respect to the Christoffel symbols $\big\{^{\,\gamma} _{\nu\mu}\big\}$. The above expression can be re-written in the following form
\bea\label{EXCREW}
K_{ab}&\equiv&\left[\f{\partial n_{\nu}}{\partial X^{\mu}}-\big\{^{\,\gamma} _{\nu\mu}\big\}n_{\gamma}\right]e_{a}^{\nu}e_{b}^{\mu}
=\f{\partial n_{\nu}}{\partial X^{\mu}}e_{a}^{\nu}e_{b}^{\mu}-\big\{^{\,\gamma} _{\nu\mu}\big\}e_{a}^{\nu}e_{b}^{\mu}n_{\gamma}\nonumber\\
&=&\f{\partial}{\partial X^{\mu}}\left(n_{\nu}e^{\nu}_{a}\right)e^{\mu}_{b}-\f{\partial}{\partial X^{\mu}}\left(\f{\partial X^{\nu}}{\partial y^a}\right)\f{\partial X^{\mu}}{\partial y^b}n_{\nu}
-\big\{^{\,\gamma} _{\nu\mu}\big\}\f{\partial X^{\nu}}{\partial y^a}\f{\partial X^{\mu}}{\partial y^b}n_{\gamma}\nonumber\\
&=&-\f{\partial^2 X^{\nu}}{\partial y^a \partial y^c}\f{\partial y^c}{\partial X^{\mu}}\f{\partial X^{\mu}}{\partial y^b}n_{\nu}-\big\{^{\,\gamma} _{\nu\mu}\big\}\f{\partial X^{\nu}}{\partial y^a}\f{\partial X^{\mu}}{\partial y^b}n_{\gamma}\nonumber\\
&=&-n_{\nu}\left[\f{\partial^2 X^{\nu}}{\partial y^a \partial y^b}-\big\{^{\,\nu} _{\alpha\mu}\big\}\f{\partial X^{\alpha}}{\partial y^a}\f{\partial X^{\mu}}{\partial y^b}\right].
\eea
Now, if we take $X_{+}^{\mu}(y^{a})$ and $X_{-}^{\mu}(y^{a})$ as parametric relations for the hypersurfaces $\Sigma^{+}$ and $\Sigma^{-}$ on the exterior and interior regions, we get, respectively
\be\label{EXC}
K^{\pm}_{ab}=-n_{\mu}^{\pm}\left[\frac{\partial^2X_{\pm}^{\mu}}{\partial y^a\partial y^b}+\hat{\Gamma}^{\mu\pm}_{\nu\sigma}\frac{\partial X_{\pm}^{\nu}}{\partial y^a}\f{\partial X_{\pm}^{\sigma}}{\partial y^b}\right].
\ee
Here we should pay attention to the second term containing the asymmetric affine connection defined in (\ref{TEQ}). Since in EC theory torsion cannot propagate outside the spin matter distribution (it is non-vanishing only inside the matter) \cite{VDSabbata-book}, the connection for the exterior region is precisely the Christoffel symbol whose non-vanishing components are
\bea\label{CHREX}
\big\{^{\,v} _{vv}\big\}^{+}=-\frac{ f_{,{\mathcal R}}}{2},~~~\big\{^{\,{\mathcal R}} _{vv}\big\}^{+}=\frac{1}{2}\left( f_{,v}+ff_{,{\mathcal R}}\right),~~~
\big\{^{\,{\mathcal R}} _{v{\mathcal R}}\big\}^{+}=\frac{ f_{,{\mathcal R}}}{2},
\eea
where \lq\lq{}$,_{\mathcal R}\equiv\partial/\partial{\mathcal R}$\rq\rq{}, \lq\lq{}$,_v\equiv\partial/\partial v$\rq\rq{} and $f=f(v,{\mathcal R})$.
In order to calculate the nonzero components of (\ref{EXC}) we proceed by noting that
\bea\label{NoteP}
\frac{\partial^2X_{+}^{v}}{\partial\tau^2}&=&\f{1}{2\left(f\dot{v}^2+2\dot{{\mathcal R}\dot{v}}\right)^2}\bigg[2\ddot{v}\left(f\dot{v}^2+2\dot{{\mathcal R}\dot{v}}\right)
-\dot{v}\left(\dot{{\mathcal R}}f_{,{\mathcal R}}\dot{v}^2+2f\dot{v}\ddot{v}+2\ddot{{\mathcal R}}\dot{v}+2\dot{{\mathcal R}\ddot{v}}\right)\bigg],\nn
\frac{\partial^2X_{+}^{{\mathcal R}}}{\partial\tau^2}&=&\f{1}{2\left(f\dot{v}^2+2\dot{{\mathcal R}\dot{v}}\right)^2}\bigg[2\ddot{{\mathcal R}}\left(f\dot{v}^2+2\dot{{\mathcal R}\dot{v}}\right)
-\dot{{\mathcal R}}\left(\dot{{\mathcal R}}f_{,{\mathcal R}}\dot{v}^2+2f\dot{v}\ddot{v}+2\ddot{{\mathcal R}}\dot{v}+2\dot{{\mathcal R}\ddot{v}}\right)\bigg].
\eea
Substituting the above expressions into equation (\ref{EXC}) and noting that $\partial\theta/\partial\tau=\partial\phi/\partial\tau=0$, we finally get
\bea\label{EXCEX}
K_{tt}^{+}=-\frac{\dot{v}^2\left[ff_{,{\mathcal R}}\dot{v}+f_{,v}\dot{v}+3f_{,{\mathcal R}}\dot{{\mathcal R}}\right]+2\left(\dot{v}\ddot{{\mathcal R}}-\dot{{\mathcal R}}\ddot{v}\right)}{2\left(f\dot{v}^2+2\dot{{\mathcal R}\dot{v}}\right)^{\frac{3}{2}}},~~~~~~~
K^{+\theta}_{\theta}&=&K^{+\phi}_{\phi}=\frac{f\dot{v}+\dot{{\mathcal R}}}{{\mathcal R}\sqrt{f\dot{v}^2+2\dot{{\mathcal R}}\dot{v}}}.
\eea
In the process of obtaining the extrinsic curvature of $\Sigma$, proceeding from the interior region, we should note that the connection through which the extrinsic curvature tensor is calculated is no longer symmetric owing to the presence of the spin matter. Therefore, we need to begin with
\bea\label{ACON}
\hat{\Gamma}^{\mu-}_{~\nu\alpha}=\hat{\Gamma}^{\mu-}_{~(\nu\alpha)}+\hat{\Gamma}^{\mu-}_{~[\nu\alpha]}
=\big\{^{\,\mu}_{\nu\alpha}\big\}^{-}+ {K}^{\mu}_{~\nu\alpha}
=\big\{^{\,\mu} _{\nu\alpha}\big\}^{-}+\f{1}{2}\left(T^{\mu}_{~\nu\alpha}-T_{\nu~\alpha}^{~\mu}-T_{\alpha~\nu}^{~\mu}\right).
\eea
Substituting for the torsion tensor from equation (\ref{Tconstraint}) and rearranging the terms we find
\bea\label{CONSPIN}
\hat{\Gamma}^{\mu-}_{~\nu\alpha}=\big\{^{\,\mu} _{\nu\alpha}\big\}^{-}+
\f{\kappa^2}{2}\bigg[2\big(\epsilon_{\nu~\alpha}^{~\mu~\rho}+\epsilon_{\alpha~\nu}^{~\mu~\rho}-
\epsilon^{\mu~~\rho}_{~\nu\alpha}\big)J_{\rho}+\f{1}{2}\big(S_{\nu\alpha}u^{\mu}-S^{\mu}_{~\alpha}u_{\nu}-
S^{\mu}_{~\nu}u_{\alpha}\big)\bigg].
\eea
A subsequently suitable spacetime averaging reveals that the second term in square brackets vanishes since the connection is linear with respect to the spin density tensor. However, the first term may not generally become zero since the axial current is a timelike vector field. Substituting  the averaged affine connection in the minus sign of (\ref{EXC}) we have
\bea\label{EXCAFF}
K^{-}_{ab}=-n_{\mu}^{-}\bigg[\frac{\partial^2X_{-}^{\mu}}{\partial y^a\partial y^b}+\big\{^{\,\mu} _{\nu\alpha}\big\}^{-}\frac{\partial X_{-}^{\nu}}{\partial y^a}\f{\partial X_{-}^{\alpha}}{\partial y^b}+\kappa^2\big(\epsilon_{\nu~\alpha}^{~\mu~\rho}+\epsilon_{\alpha~\nu}^{~\mu~\rho}-\epsilon^{\mu~~\rho}_{~\nu\alpha}\big)\frac{\partial X_{-}^{\nu}}{\partial y^a}\f{\partial X_{-}^{\alpha}}{\partial y^b}\langle J_{\rho}\rangle\bigg],
\eea
from which we readily find that the third term in parentheses vanishes due to antisymmetrization property of the Levi-Civita tensor and  partial derivatives. After a straightforward calculation we find
\bea\label{EXCIN}
K_{tt}^{-}=0,~~~~~~K^{-\theta}_{\theta}=K^{-\phi}_{\phi}=\frac{1}{r_{\Sigma}a(t)}.
\eea
However, due to the presence of the third term in (\ref{EXCAFF}) there may remain other components of the extrinsic curvature tensor though the spacetime is spherically symmetric. Let us calculate them to show that these terms vanish too. The $(t,\theta)$ component reads
\bea\label{TTEXC}
K^{-}_{t\theta}=-a(t)\bigg\{\frac{\partial^2r}{\partial t \partial\theta}+\big\{^{~ r} _{\nu\alpha}\big\}^{-}\f{\partial X_{-}^{\nu}}{\partial t}\f{\partial X_{-}^{\alpha}}{\partial \theta}+\kappa^2\big(\epsilon_{\nu~\alpha}^{~r~\rho}+\epsilon_{\alpha~\nu}^{~r~\rho}\big)\f{\partial X_{-}^{\nu}}{\partial t}\f{\partial X_{-}^{\alpha}}{\partial \theta}\langle J_{\rho}\rangle\bigg\}.
\eea
Since $\langle J_{\rho}\rangle$ has only a time component, the Levi-Civita tensor vanishes  and we get
\be\label{KTTHETA}
K^{-}_{t\theta}=K^{-}_{t\phi}=0.
\ee
The remaining components can be calculated in the same way as
\be\label{KTHPHI}
K^{-}_{\theta\phi}=K^{-}_{\phi\theta}=-\kappa^2a(t)\big(\epsilon_{\theta~\phi}^{~r~t}+\epsilon_{\phi~\theta}^{~r~t}\big)\langle J_{t}\rangle=0,
\ee
since both the Levi-Civita tensors are equal but with opposite signs.
Using (\ref{FFF1}), the continuity of the extrinsic curvatures across $\Sigma$ implies the following relations
\bea
f\dot{v}+\dot{{\mathcal R}}&=&1,\label{A15}\\
\dot{v}^2\left[(ff_{,{\mathcal R}}+f_{,v})\dot{v}+3f_{,{\mathcal R}}\dot{{\mathcal R}}\right]&+&2\left(\dot{v}\ddot{{\mathcal R}}-\dot{{\mathcal R}}\ddot{v}\right)=0.\nn\label{A16}
\eea
Taking derivatives of (\ref{A15}) and the first part of (\ref{FFF1}) we have
\be\label{A17}
2\dot{{\mathcal R}}\ddot{v}=\dot{f}\dot{v}^2,~~~~~~2\dot{{\mathcal R}}\ddot{{\mathcal R}}=-\dot{f}\dot{v}\left(2\dot{{\mathcal R}}+f\dot{v}\right),
\ee
hence we can construct the following relation as
\be\label{REC}
\ddot{{\mathcal R}}\dot{v}=-\ddot{v}\left(2\dot{{\mathcal R}}+f\dot{v}\right).
\ee
Substituting (\ref{REC}) and the first part of (\ref{A17}) into (\ref{A16}), we finally find
\be\label{KTTPF}
K_{tt}^{+}=-\frac{f_{,v}\dot{v}^2}{2\dot{{\mathcal R}}}=0,
\ee
which clearly shows that $f(v,{\mathcal R})$ must be a function of ${\mathcal R}$ only. Solving equations (\ref{A15}) and the first part of (\ref{FFF1}) we get the four-velocity of the boundary, as seen from an exterior observer
\be\label{4V}
V^{\alpha}=\left(\dot{v},\dot{{\mathcal R}},0,0\right)=\left[\f{1+\sqrt{1-f}}{f},-\sqrt{1-f},0,0\right],
\ee
where a minus sign for $\dot{{\mathcal R}}$ has been chosen since we are dealing with a collapse setting. From the second component of the above vector field and the interior solution (\ref{HH}), we find for a smooth matching of the interior and exterior spacetimes that, ${\mathcal M}({\mathcal R},v)=m(t,r_{\Sigma})$. We thus have
\be\label{INMOUTM}
2{\mathcal M}({\mathcal R},v)=\f{\Lambda}{3}{\mathcal R}^3+2{\mathcal M}_{0}{\mathcal R}^{-3w}+2{\mathcal S}_0{\mathcal R}^{3+n},
\ee
where ${\mathcal M}_0=(1/12)\kappa^2C_0r_{\Sigma}^{3(1+w)}$ and ${\mathcal S}_0=(1/2)\ell r_{\Sigma}^{-n}$. Here the new  term ${\mathcal S}_0$ is treated as a correction introduced by the spin contribution. Therefore, the line element for exterior spacetime reads
\bea\label{EXSP}
ds_{+}^2=\left[1-\f{\Lambda}{3}{\mathcal R}^2-2{\mathcal M}_{0}{\mathcal R}^{-(1+3w)}-2{\mathcal S}_0{\mathcal R}^{n+2}\right]dv^2
+2dvd{\mathcal R}-{\mathcal R}^2(d\theta^2+\sin^2\theta d\phi^2).
\eea

The location of the apparent horizon is marked by requiring that $2{\mathcal M}={\mathcal R}$ which lies on the boundary surface $r=r_{\Sigma}$ if $m=R$, or simply from equation (\ref{2MR}), $r_{\Sigma}^2\dot{a}^2=1$. Thus, we find that once the collapse velocity satisfies the following equation
\be\label{HEQ}
|\dot{a}|=\f{1}{r_{\Sigma}},
\ee
a dynamical horizon forms intersecting the boundary. Then, if the collapse velocity is bounded the boundary surface can be chosen so that no horizon forms \cite{R1}.

Now, from the first part of equation (\ref{HH}), we see that for specific values of $n$ and $w$, taken from the brown region of figure \ref{Fig1}, the collapse velocity tends to infinity. Thus, there is no minimum value for $r_{\Sigma}$ (or correspondingly a minimum mass for the collapsing volume) so that the horizon can be avoided. In contrast, for $n$ and $w$ taken from the yellow region of figure \ref{Fig1}, the speed of collapse stays bounded until the singularity time at which the scale factor vanishes. This means that to satisfy the horizon condition in the limit of approach to the singularity, the boundary of the volume must be taken at infinity which is physically irrelevant. Thus we can always take the surface boundary so that the apparent horizon is avoided. Therefore, if the collapse velocity is bounded we can take the boundary surface to be sufficiently small so that the formation of horizon is avoided during the entire phase of contraction. Furthermore, the null geodesic that has just escaped from the outermost layer of the mass distribution of the cloud ($r_{\Sigma},t_s$) can be extended to the exterior region exposing the singularity to external observers. For bouncing solutions, as the right panel of figure \ref{abra} shows, $|\dot{a}|$ remains finite throughout the collapsing and expanding phases, thus by a suitable choice of the boundary surface, the apparent horizon is failed to cover the bounce.

For $n=-3$, $w=0$ and $\sigma_0^2=48J_0^2$ the exterior spacetime metric is written as
\bea\label{SchAds}
ds_{+}^2=\left[1-\f{\Lambda}{3}{\mathcal R}^2-\f{2{\mathcal M}^{\prime}_{0}}{{\mathcal R}}-\f{2{\mathcal S}^{\prime}_0}{\mathcal R}\right]dv^2
+2dvd{\mathcal R}-{\mathcal R}^2(d\theta^2+\sin^2\theta d\phi^2),
\eea
where ${\mathcal M}^{\prime}_{0}=(1/12)\kappa^2\rho_i a_i^3 r_{\Sigma}^3$ and ${\mathcal S}^{\prime}_0=(-1/48)\kappa^4\sigma_0^2 r_{\Sigma}^3$, clearly exhibiting a
Schwarzchild-anti-de Sitter metric in retarded null coordinates \cite{SCHNDE} with a constant mass ${\rm M}_0={\mathcal M}^{\prime}_0+{\mathcal S}^{\prime}_0$  with corrections due to spin contributions.
\section{concluding remarks}\label{CON}

The study of the end-state of matter gravitationally
collapsing becomes quite interesting
when averaged spin degrees of freedom and torsion are taken into account. To our knowledge, the literature
concerning this line of research is somewhat scarce\footnote{With respect to the initial cosmological singularity, there seems to have been more efforts in analyzing it when fermionic terms impose modifications to the classical equations (explicitly by means of fermionic degrees of freedom being  present or induced by means of some averaged quantities), see e.g., (\cite{GAS,GEA,Gasprini,Shapiro,obuk,Nurga,AVCS,Pop,Teixeira}).}, see e.g., \cite{match}. Torsion is perhaps one of the important consequences of coupling gravity to
fermions. In general, this leads to non-Riemannian
spacetimes where departures from the dynamics of GR would be
expected and should be explored. The well known and established CSK \cite{Ortin:2004ms} theories can also be a starting point\footnote{A collapse setting was introduced in \cite{MALFERG} where the non-minimal coupling of classical gravity to fermions results in the singularity avoidance}. Nevertheless, the explicit presence of fermionic fields may not provide a simple enough set-up to investigate the final outcome of a
gravitational collapse. There are, however, other, perhaps more
manageable scenarios. They employ torsion just to mimic the effects
of matter with spin degrees of freedom on gravitational systems.

It was in that precise context that we have therefore considered the approach presented in this  paper. More precisely, we studied the gravitational collapse of a cloud whose  matter content
was taken as a Weyssenhoff fluid \cite{Weys} in the context of the
EC theory \cite{Hehl-Heyde}, i.e., with torsion. A negative $\Lambda$ was included to provide
an initially positive pressure, so that a collapse process could initially be set up. The torsion is not, however,
a dynamical field, allowing it to be eliminated in favor of algebraic
expressions.

In addition, we have restricted ourselves to a special but manageable spacetime model where the interior region line element is a FLRW metric, allowing a particularly manageable framework to investigate. The corresponding effective energy momentum from a macroscopic perspective has a
perfect fluid contribution plus those  induced from averaged spin
interactions. A relevant feature is that this effective matter can, within specific conditions, convey a negative pressure  effect. As a consequence, this may induce the avoidance of the formation of trapped
surfaces, from one hand, and the possibility of singularity removal from the other hand.

In a compact manner, our main results are as follows:

\begin{itemize}

\item For singular solutions ($\ell>0$), the formation or otherwise of trapped surfaces not only depend on the equation of state parameter but also on the spin density divergence term ($n$). Therefore, from determining the initial setting subject to ($i$) the regularity condition on the absence of trapped
surfaces at an initial epoch, ($ii$) the validity of the energy
conditions and ($iii$) the positivity of the effective pressure at
an initial time, trapped surfaces can either develop (for $n<-2, w>-1/3$) or be avoided ($-2<n<0, w<-1/3$) throughout the collapse.

\item A special case in which the equation of state of spin fluid
is $p_{{\rm SF}}=-\rho_{{\rm SF}}$ was considered separately and it was found that no singularity occurs. This very unorthodox case can be thought of as a stationary state.
\item The set of collapse solutions can be categorized through the six-dimensional space of the parameters ($J_i^2,\sigma_i^2,n,w,a_i,\rho_{i_{{\rm SF}}}$) so that the first two  are related to initial values for spin source parameters (note that $J_i^2=J_0^2a_i^n$ and $\sigma_i^2=\sigma_0^2a_i^n$). The next two parameters are the rate of divergence of spin density and barotropic index and the last two  are the initial values of the scale factor and energy density. Each point from this space represents a collapse process that can be either led to a spacetime singularity or a non-singular bounce. Determining the suitable ranges for this set of initial data is not straightforward and so, for the sake of clarity, we have to deal with the two-dimensional subspaces by fixing four of the above parameters. However, we could infer that among the allowed sets of the initial data we can always pick up those for which trapped surfaces are prevented (in singular solutions) during the collapse scenario, (see the regions in figure \ref{Fig1}), where we have fixed the same initial values for energy density and scale factor.

\item Depending on the initial value of energy density and the source parameters related to spin-spin contact interaction and axial current, singular ($\ell>0$ and $C_0>0$) and non-singular ($\ell<0$ and $C_0>0$) solutions can be found. In the former the singularity occurs sooner than the case in which the spin correction term is neglected (see the left panel in figure \ref{sftb}). For the non-singular scenario, the collapse process halts at a finite value of the scale factor and then turns to expansion. 
\end{itemize}

It is worth mentioning that besides the model presented here,
non-singular scenarios have been reported within models of $f(R)$ theories of gravity in Palatini \cite{PAL} and metric \cite{METFRB} formalisms, generalized teleparallel theories of gravity \cite{GTG}, bouncing in brane models \cite{BBM} and modified Gauss-Bonnet gravity \cite{MGB} (see also \cite{REPBC} for recent review). In \cite{PARKFULL}, it is shown that a quantized neutral scalar field minimally coupled to classical gravitational field may avoid the singularity.

It is also interesting to note that, beside the Frenkel condition we employed here, we could consider the possibility of relaxing it, therefore allowing to take a more general matter content. If such a modification is employed, then the number of degrees of freedom of the torsion tensor would increase, seemingly bringing a
more complicated setting to deal with.

Finally, we would like to present a few possible additional subsequent lines of exploration.

Although being a wider set-up with respect to GR, it could be fruitful to generalize
action (\ref{action}). More concretely, replacing the cosmological constant  by some scalar matter. This would allow for the
establishment of limits for the dominance of any matter component
(and associated intrinsic effects) towards a concrete gravitational
collapse outcome where, for example, bosonic and fermionic matter  would be competing. Perhaps more challenging would be  to
employ a Weyssenhoff fluid description that could have different
features whether we use $s=\frac{1}{2}$ fermion or a
Rarita-Schwinger field with $s=\frac{3}{2}$ spin angular momentum.
The gravitational theory of such latter particles in the presence of
torsion has been discussed in \cite{s32}. 

\section*{Acknowledgments}

The authors are grateful to D. Malafarina and R. Goswami for useful discussions and to F. W. Hehl for helpful correspondence. A. Ranjbar also would like to thank F. Canfora and J. Zanelli for their interesting comments.
\appendix
\section{On the Einstein-Cartan action}
The Riemann tensor in EC theory reads
\bea\label{ECRI}
\hat{R}^{\lambda}\!\!~_{\mu\nu\rho}=\partial_{\nu}\hat{\Gamma}^{\lambda}\!\!~_{\mu\rho}-
\partial_{\rho}\hat{\Gamma}^{\lambda}\!\!~_{\mu\nu}+\hat{\Gamma}^{\sigma}\!\!~_{\mu\rho}
\hat{\Gamma}^{\lambda}\!\!~_{\sigma\nu}-\hat{\Gamma}^{\sigma}\!\!~_{\mu\nu}
\hat{\Gamma}^{\lambda}\!\!~_{\sigma\rho}.
\eea
The general affine connection can be written in terms of Christoffel symbols and contorsion (we note that in EC theory $\hat{\nabla}_{\alpha}g_{\mu\nu}=0$)
\be\label{GAFC}
\hat{\Gamma}^{\alpha}\!\!~_{\beta\gamma}=\big\{^{\alpha} _{\beta\gamma}\big\}+K^{\alpha}\!\!~_{\beta\gamma}.
\ee
Substituting the above expression into (\ref{ECRI}) we get
\bea\label{ECRI1}
\hat{R}^{\lambda}\!\!~_{\mu\nu\rho}&=&R^{\lambda}\!\!~_{\mu\nu\rho}(\{\})+\partial_{\nu}K^{\lambda}\!\!~_{\mu\rho}-\partial_{\rho}K^{\lambda}\!\!~_{\mu\nu}
+\big\{^{\sigma} _{\mu\rho}\big\}K^{\lambda}\!\!~_{\sigma\nu}+\big\{^{\lambda} _{\sigma\nu}\big\}K^{\sigma}\!\!~_{\mu\rho}+K^{\sigma}\!\!~_{\mu\rho}K^{\lambda}\!\!~_{\sigma\nu}\nn
&-&\big\{^{\sigma} _{\mu\nu}\big\}K^{\lambda}\!\!~_{\sigma\rho}-\big\{^{\lambda} _{\sigma\rho}\big\}K^{\sigma}\!\!~_{\mu\nu}-K^{\sigma}\!\!~_{\mu\nu}K^{\lambda}\!\!~_{\sigma\rho},
\eea
whereby contracting twice gives the Ricci scalar as
\bea\label{RICCI}
\hat{R}&=&R(\{\})+g^{\mu\rho}\partial_{\nu}K^{\nu}\!\!~_{\mu\rho}-g^{\mu\rho}\partial_{\rho}K^{\nu}\!\!~_{\mu\nu}+g^{\mu\rho}\big\{^{\sigma} _{\mu\rho}\big\}K^{\nu}\!\!~_{\sigma\nu}
+\big\{^{\lambda} _{\sigma\lambda}\big\}K^{\sigma\mu}\!\!~_{\mu}+K^{\sigma\mu}\!\!~_{\mu}K^{\lambda}\!\!~_{\sigma\lambda}-g^{\mu\rho}\big\{^{\sigma} _{\mu\nu}\big\}K^{\nu}\!\!~_{\sigma\rho}\nn
&-&g^{\mu\rho}\big\{^{\nu} _{\sigma\rho}\big\}K^{\sigma}\!\!~_{\mu\nu}-g^{\mu\rho}K^{\sigma}\!\!~_{\mu\nu}K^{\nu}\!\!~_{\sigma\rho}.
\eea
The metricity condition ($\nabla_{\alpha}g_{\mu\nu}=0$) leaves us with the following expressions
\bea\label{PMET}
\partial_{\nu}g^{\mu\rho}=-g^{\alpha\rho}\big\{^{\mu} _{\nu\alpha}\big\}-g^{\mu\alpha}\big\{^{\rho} _{\nu\alpha}\big\},~~~~
\partial_{\rho}g^{\mu\rho}=-g^{\alpha\rho}\big\{^{\mu} _{\rho\alpha}\big\}-g^{\mu\alpha}\big\{^{\rho} _{\rho\alpha}\big\},
\eea
by the virtue of which we can simplify (\ref{RICCI}) to finally get
\bea\label{RICCIF}
\hat{R}=R(\{\})+\nabla_{\lambda}K^{\lambda\rho}\!\!~_{\rho}-\nabla_{\rho}K^{\lambda\rho}\!\!~_{\lambda}+K^{\sigma\mu}\!\!~_{\mu}K^{\lambda}\!\!~_{\sigma\lambda}
-K^{\sigma\rho}\!\!~_{\nu}K^{\nu}\!\!~_{\sigma\rho}.
\eea
Performing the integration and neglecting the total derivatives we arrive at the integrand given in the expression (\ref{action}).

\end{document}